# INTEGRATION AND CONTAGION IN US HOUSING MARKETS

John Cotter[1], Stuart Gabriel[2] and Richard Roll[3]

## ABSTRACT

This paper explores integration and contagion among US metropolitan housing markets. The analysis applies Federal Housing Finance Agency (FHFA) house price repeat sales indexes from 384 metropolitan areas to estimate a multi-factor model of U.S. housing market integration. It then identifies statistical jumps in metropolitan house price returns as well as MSA contemporaneous and lagged jump correlations. Finally, the paper evaluates contagion in housing markets via parametric assessment of MSA house price spatial dynamics.

A R-squared measure reveals an upward trend in MSA housing market integration over the 2000s to approximately .83 in 2010. Among California MSAs, the trend was especially pronounced, as average integration increased from about .55 in 1997 to close to .95 in 2008! The 2000s bubble period similarly was characterized by elevated incidence of statistical jumps in housing returns. Again, jump incidence and MSA jump correlations were especially high in California. Analysis of contagion among California markets indicates that house price returns in San Francisco often led those of surrounding communities; in contrast, southern California MSA house price returns appeared to move largely in lock step.

The high levels of housing market integration evidenced in the analysis suggest limited investor opportunity to diversify away MSA-specific housing risk. Further, results suggest that macro and policy shocks propagate through a large number of MSA housing markets. Research findings are relevant to all market participants, including institutional investors in MBS as well as those who regulate housing, the housing GSEs, mortgage lenders, and related financial institutions.

[1]UCD School of Business, University College Dublin, Blackrock, Co. Dublin, Ireland. Email: john.cotter@ucd.ie and Research Fellow, Ziman Center for Real Estate, UCLA Anderson School of Management.
[2]Anderson School of Management, University of California, Los Angeles, 110 Westwood Plaza, Los Angeles, California 90095, stuart.gabriel@anderson.ucla.edu
[3] Anderson School of Management, University of California, Los Angeles, 110 Westwood Plaza, Los Angeles, California 90095, rroll@anderson.ucla.edu.
The authors gratefully acknowledge research support from the UCLA Ziman Center for Real Estate. Cotter also acknowledges the support of Science Foundation Ireland under Grant Number 08/SRC/FM1389.

## I. Introduction

The boom and bust of house prices defined the opening decade of the 21st century. As reported in Shiller (http://www.econ.yale.edu/shiller/data.htm)), US national house prices recorded a decline of 31 percent over the 2006 - 2010 period, about on par with the peak-to-trough contraction during the Great Depression. Implosion in house prices figured importantly in the 2007 meltdown in mortgage and capital markets and the downturn in the global economy. As shown in Figure 1, the fall-off in house prices and related economic decline were especially severe in California.

Neither analysts on Wall St., regulators in Washington, D.C., nor most academic economists anticipated the magnitude of the house price cycle, its geographic reach, or related housing market contagion. Indeed, while urban economists long have addressed linkages among a system of cities (see, for example, Henderson (1977)), few studies have focused on the spatial-temporal structure of the house price cycle. For example, little is known about the relative exposure of MSA housing markets to fluctuations in the national economy, despite the importance of such to investor diversification or to economic policy propagation. Also, few analyses have provided insights as regards the metropolitan geography of house price returns, notably including incidence of extreme (jump) returns and the directions of contagion.

We address those questions via analyses of integration, correlation, and contagion in US metropolitan housing markets. Those estimates are important to investors as they provide an indication of opportunities to diversify away metropolitan-specific housing risk. For example, high levels of MSA integration and contagion among geographically distinct residential markets could mitigate the efficacy of geographic diversification strategies implemented by investors in mortgage-backed securities. An improved understanding of metropolitan housing market integration also could provide new insights regarding the spatial incidence of national economic policy. In general, measures of integration and contagion in housing markets provide signals for price return performance and are relevant for the full spectrum of market participants, be they lenders, housing and mortgage investors, homebuilders, and the like.

Following Pukthuanthong-Le and Roll (2009), we compute a simple intuitive measure of housing market integration, based on the proportion of an MSA's housing market returns that can be explained by an identical set of national factors. The level of integration is associated with the magnitude of R-Square, with higher values indicating higher levels of integration. Two MSAs are viewed as perfectly integrated if those same national factors fully explain housing market returns in both those areas. In that case, there would be a R-square of 1 so there is no diversification potential between the MSAs.

Results of the analysis indicate elevated and increasing MSA housing market integration. For the US as a whole, housing market integration trended up over the decade of the 2000s to about .83 in 2010.[1] In California the trend was marked; there average housing market integration moved up from about .55 in 1997 to close to .95 in 2008! Also noteworthy, however, was the abrupt downward adjustment in California integration, to approximately .75, in the wake of the recent severe implosion in house prices. Further disaggregation of California trends revealed more pronounced declines in integration among coastal markets in the context of the housing bust. That result likely reflects special factors associated with coastal markets (supply constraint, presence of amenities, and lack of subprime lending) in the context of ongoing weakness in national economic and housing market fundamentals.

Using the Lee and Mykland (2008) measure to characterize extreme returns, we find that the 2000s bubble period also was distinguished by a relatively high incidence of jumps in housing returns. Jumps were especially evident early in the boom during 2004-2005 as well as in 2008 in the wake of the bust in house prices, the latter likely owing to extreme declines in returns in certain MSAs. During early stages of the boom (2003 – 2004), return jumps in California suddenly become very prevalent with close to 70 percent of cities having significant extreme housing returns; further, during that period, the jumps were ubiquitous among coastal and inland California cities. In marked contrast, during the 2007-2008 bust and among California MSAs, only inland cities witnessed extreme movements in housing returns. Inland cities are characterized by a lack of constraint on housing supply and, in hindsight, they had been substantially overbuilt. Further, those areas had been the focus of substantial boom period subprime lending. As boom turned to bust, inland areas of California quickly and largely imploded.

As would be expected, both in the US overall and in California, metropolitan return correlations are dramatically larger than jump return correlations in both incidence and magnitude. California, however, stands apart from the rest of the US in both returns and extreme returns. Research findings indicate relatively high levels of housing return and jump return correlations in California compared to the rest of the US. For example, contemporaneous housing return correlations are generally in the range of 0.2 – 0.3 with about 20 percent significant for MSAs outside California. In marked contrast, in excess of 92 percent of California MSA returns were significantly correlated with a mean correlation level of about .66! Similar results are obtained for lead (one quarter ahead) MSA correlations. Among areas outside California, less than 10 percent of lead correlations were statistically significant with mean lead correlation levels at or below 0.20. In California, more than three-quarters of MSAs recorded significant lead return correlations with a mean correlation level of about .57.

California also was markedly different as regards contemporaneous and lead LM jump correlations. Among areas outside of California, significant contemporaneous jump correlations were small in number and in the range of only .02 – .03. Large lead jump correlations outside California similarly occurred infrequently in any census division with mean correlation coefficients (except for New England) of .04 or less. In contrast, both the incidence and magnitude of contemporaneous and lead jump correlations were greater for California.

[1] A measure of 1.0 would indicate perfectly integrated markets while zero would indicate no integration at all; hence, the observed average of 0.83 implies that U.S. housing markets are 83% integrated relative to the maximum possible level.

Given the above aberrant nature of integration, jump incidence, and MSA jump correlations among California MSAs, the analysis turns to parametric assessment of spatial and temporal contagion among California cities. Regression analyses over the full sample timeframe indicate that house price returns for Los Angeles and surrounding areas largely move in lock-step. In contrast, findings for Bay Area regional housing markets provide some evidence of a spatial term structure of contagion. In that region, housing returns in San Francisco lead those of many northern California communities. Contagion findings are robust to controls for booms and busts in California housing markets.

The plan of the paper is as follows. Section II provides assessment of integration of US MSA house price returns. In Section III, we report on analyses of both contemporaneous and lagged correlations and jump correlations in MSA house price returns. Section IV provides tests of geographic-temporal contagion among MSA housing markets in California. In section V, we provide concluding remarks.

## II. Integration

Substantial research has been undertaken as regards integration of international equity markets. The applications vary in geography of focus, as some papers address integration in the European community (see, for example, Hardouvelis, Malliaropoulos, and Priestley (2006), and Schotman and Zalewska (2006)), whereas others investigate emerging markets (see, for example, Bakaert and Harvey (1995), Chamber and Gibson (2006), Bakaert, Harvey, Lundblad and Siegel (2008)). The analyses also vary widely in methodological approach. For instance, Carrieri, Errunza and Hogan (2007) use GARCH-in-mean methods to assess correlation in returns and volatility between markets, whereas Longin and Solnik (1995) use cointegration techniques. While integration is often described in terms of cross-country correlations in stock returns (for an early study see King and Wadhwani (1990)), such a measure is argued to be flawed. Indeed, in the case where multiple factors drive returns, markets may be imperfectly correlated but perfectly integrated.[2]

[2] As shown by Pukthuanthong and Roll (2009), while perfect integration implies that identical global factors fully explain index returns across countries, some countries may differ in their sensitivities to those factors and accordingly not exhibit perfect correlation. An easy intuitive example would be an energy-exporting country such as Saudi Arabia and an energy-importing country such as Hong Kong. Both countries might be positively associated with global factors such as consumer goods or financial services. Morever, both countries could be fully integrated in the global economy; yet the simple correlation between their stock market returns could be relatively small, or even negative, because higher energy price increase Saudi equity values and decrease Hong Kong equity values. As a

As suggested by Pukthuanthong-Le and Roll (2009), a simple intuitive measure of financial market integration is the proportion of a country's returns that can be explained by an identical set of [global] factors. This measure of integration focuses on the magnitude of country-specific residual variance in a factor model seeking to explain a broadly-defined country equity return index.[3] Clearly, to the extent global factors explain only a small proportion of variance in a country's returns, the country would be viewed as less integrated (see, for example, Stulz (1981) and Errunza and Losq (1985)).[4] In contrast, markets would be viewed as highly integrated to the extent their returns are explained. We below describe US metropolitan housing markets as highly integrated if identical US national factors explain a large portion of the variance in MSA house price returns. To compute US housing market integration, we regress metropolitan house price returns on an identical set of national economic and housing market fundamentals.

Integration is viewed as important to investors, policymakers, and market participants in general. A measure of housing market integration provides some indication of the benefits to investor diversification among MSA markets. While there may be some benefit to diversifying away MSA-specific housing market risk, those benefits would decline with increases in integration. Indeed, high levels of integration may mitigate strategies of geographic diversification among investors in mortgage-backed securities. Also, among other things, a measure of metro housing market integration would provide national economic policymakers with some indication of the geographic ubiquity of policy propagation. High levels of MSA housing return integration imply that those markets largely are driven by national factors, notably including monetary policy and other housing fundamentals. Similarly, elevated levels of metro housing market integration imply that macro and financial shocks will propagate through a larger number of MSA housing markets. This will have relevance for all market participants, including institutional investors in residential MBS as well as those who regulate housing, the housing GSEs, mortgage lenders, and related financial institutions.

### a. Model Specification and Data

MSA-specific house price returns are computed using the U.S. Federal Housing Finance Agency (FHFA) metropolitan indices, previously known as the OFHEO house price series.

---

consequence, the extent to which the multi-factors drive returns is a better indication of likely diversification benefits than a correlation measure.

[3] In contrast, in the presence of multiple national factors, the simple correlation between MSA house price return indexes could be a flawed measure of integration unless those MSAs have identical exposure to the national factors, e.g., unless the estimated coefficient vectors are exactly proportional across MSAs.

[4] According to this definition, a country is perfectly integrated if the country-specific variance is zero after controlling for global factors. In the case of two perfectly integrated countries, market indexes would have zero residual variance. See Pukthuanthong and Roll (2009) for discussion and details.

The FHFA series are weighted repeat-sale price indices associated with single-family homes. Home sales and refinancing activity included in the FHFA sample derive from conventional home purchase mortgage loans conforming to the underwriting requirements of the housing Government Sponsored Enterprises—the Federal National Mortgage Association (Fannie Mae) and the Federal Home Loan Mortgage Corporation (Freddie Mac). The FHFA data comprise the most extensive cross-sectional and time-series set of quality-adjusted house price indices available in the United States.[5] We compute house price returns for each MSA in our sample as the log quarterly difference in its repeat home sales price index.[6] The MSA level data are quarterly from 1975:Q1 – 2010:Q1. The number of MSAs in the database increases over time from 2 in 1975 to 380 by 1993. By the end of the sample timeframe, there are 384 MSAs in the dataset.

Per above, for each MSA in the sample, log percent change in the MSA-specific house price indices is regressed on a common set of national economic, financial and housing market factors. The specific factors and their definitions are displayed in Appendix Table 1. The factors include measures of change in population, payroll employment, unemployment rate, S&P500, industrial production, CPI, and PPI materials prices as well as personal income, consumer sentiment, single-family building permits, Fed Funds rate, 10-year constant maturity Treasury yields, and the like. All factor data are quarterly in frequency from 1975:Q1 – 2010:Q1 with the exception of consumer sentiment, which is available from 1977:Q4. Data for the factors are obtained from the Federal Reserve Bank of St. Louis FRED (Federal Reserve Economic Data) with the exception of the S&P500 (Datastream) and personal income (US Department of Commerce National Income and Product Accounts). The MSA returns series are pre-whitened to remove serial correlation. A VAR(1) is employed based on optimal AIC/BIC criteria from running the factor model on each individual MSA. The average level of integration is measured by the R-squares from the multi-factor model fitted for a 20-quarter moving window for the samples of MSAs (the use of other window sizes gave the same qualitative results). The R-squares in these moving windows indicate the corresponding levels of housing market integration.

### b. Return Regressions on National Factors

Estimation results indicate that U.S. MSA housing market integration has increased over time. Figure 2 provides information on trends in housing market integration for the MSAs in our sample. Panel A of Figure 2 shows that trend for the 1983:Q4 – 2009-Q4 period both for the national and California samples. Very little trend in US MSA housing market integration appeared during the decades of the 1980s and 1990s. In contrast, the 2000s provides graphic evidence of trending up in housing market integration among US MSAs, from about .70 in 2000 to approximately .83 by decade's end. In California the trend in housing market integration was even more marked moving up from about .55 in 1997 to close to .95 in 2008! Further noteworthy, however, was the abrupt downward adjustment in California housing market integration, to approximately .75, in the wake of the recent severe implosion in house prices. Indeed, localized factors associated with the California housing bust resulted in some disassociation of California metropolitan housing returns from national economic fundamentals.

[5] For a full discussion of the OFHEO house price index, see "A Comparison of House Price Measures", Mimeo, Freddie Mac, February 28, 2008.

[6] In principle, it would be desirable to model house prices at higher frequencies. Unfortunately, monthly quality-adjusted house price indices are available from OFHEO only for Census Divisions (N=18) and only for a much shorter time frame.

We control for potential bias in the FHFA data in terms of when an MSA was included in the database. Regardless, the finding of increased integration still holds. Panel B of Figure 2 shows the average R-square pattern for 3 time cohorts. This categorization of MSAs into cohorts assesses the robustness of results to the timeframe of city inclusion in the sample. In this regard, it is possible that MSAs that entered the sample later were characterized by lower or higher R-squares. If that were the case, averaging all MSAs together could move the trend in the average either up or down. We plotted trends in the average level of integration for three time-based cohorts. The cohorts included the full timeframe of 1983:Q4 – 2010:Q1 (cohort 1), 1989:Q2 – 2010:Q1 (cohort 2), and 1992:Q1 – 2010:Q1 (cohort 3). The cohorts yielded roughly similar results and indicated a longer-term trend towards MSA housing market integration. In cohort 2, for example, the average R-square moved up from about .65 in 1989 to almost .82 in 2010.

MSA housing market cross-sectional and time-series summary statistics are contained in Table 1. For the sample of MSAs, we display mean quarterly house price returns, standard deviation of returns (sigma), the R-square measure of integration, the change in R-square over the timeframe of the analysis, and the associated time trend t-statistic (R-squares for each MSA are fit to a simple linear time trend for all available quarters). Minimum values by quintile are also presented. First, it is important to note that risk and return associated with housing has been substantial. As shown, the average quarterly return for all MSA housing markets in the sample is positive at almost 1% with an average deviation of about 2.5%. Moreover, we see substantial cross sectional variation in those measures; for example, mean house price return varies from a minimum 0.43% to a sample maximum of 1.89%.

As evidenced in Table 1, the mean final period R-square of the integration model is .82, suggesting the importance of national factors in determination of MSA house price returns. The Table also indicates substantial temporal and cross-MSA variation in the integration measure. On average R-squares increase by almost 10 percent from the beginning to end of sample. In some areas, national economic and housing market fundamentals fail to explain the majority of variation in MSA-specific house price returns (min R-squared = .35) At a maximum, those same fundamentals explain a full 99 percent of variation in MSA-specific house price returns. There is also substantial variation in the change in R-squared across the sample with a standard deviation of .187. Appendix Table 2 contains integration details for all 384 MSAs.[7]

Table 2 presents integration details for the 28 California MSAs included in our dataset. Relative to the full national sample of 384 MSAs, California metropolitan areas are characterized by elevated mean house price returns, return volatility, and integration time trend t-statistic. Further discernable in Table 2 are distinct coastal versus inland housing market phenomena. Comparing coastal MSAs (see, for example, San Francisco, Oakland, San Jose, Los Angeles, Santa Ana, and Santa Barbara) with inland MSAs (for example, Bakersfield, Fresno, Madera, Merced, Modesto, Riverside, and Sacramento), note that the former are roughly characterized by relatively higher mean house price returns, lower return volatility, damped levels of integration, and lower integration trend t-statistics. Among California coastal MSAs, mean quarterly returns averaged an elevated 1.6 percent; further, integration R-squared averaged .69 with an insignificant time trend t-statistic. In marked contrast, California Central Valley and Inland Empire cities displayed substantially lower mean house price returns, elevated return volatility, higher levels of integration, and higher integration trend t-stats. In inland areas, mean quarterly house price returns were a damped 1 percent with an elevated sigma of 3.4 percent; further, the t-statistic on the integration time trend was 2.2, well in excess of t-statistics for California coastal MSAs and for the nation as a whole.

Panel C of Figure 2 shows trends in average R-square for inland and coastal MSAs in California. As is evident, average integration for MSAs in both areas trended up over the late-1990s through 2008 period. Striking is an up and down pattern in integration that roughly coincided with the boom and bust in housing markets overall. While integration levels for California MSAs moved up from about .75 to in excess of .90 in the context of the 2000s cyclical boom in housing, those same measures fell back markedly during the subsequent bust as California housing returns became increasingly divorced from national economic fundamentals. Further, the chart is suggestive that localized factors recently played a substantially greater role in determination of coastal California house price returns, as suggested in the divergence in integration between coastal and inland areas in the context of the implosion in housing markets. That divergence likely reflected special factors supportive of the performance of coastal markets (supply constraint, desirable natural amenities, shorter commutes, and the like) in the context of ongoing weakness in national economic and housing market fundamentals. As was broadly reported, Central Valley and Inland Empire cities collectively comprised the epicentre of the 2000s boom-bust cycle in California housing markets. Those areas were characterized by high levels of subprime lending, elastic land and housing supply, longer commutes, and substantial overbuilding. In many cases, the interior MSAs are outer-ring bedroom communities for employment centers closer to the coast. The results suggest distinctions in housing return phenomena both within and between California MSAs and the nation as a whole. We return to that below, in discussion of MSAs house price return correlations and contagion.

[7] The table further provides the quintile and rank (from lowest to highest) across the 384 MSAs of returns, sigma, and integration time trend t-statistic.

## III. MSA Return and Jump Return Correlations

In this section, we investigate the magnitude of metropolitan house price returns, distinguishing between common and extreme movements (jumps). Those results are benchmarked by a discussion of contemporaneous and lagged correlations in MSA house price returns. The analysis provides insights about temporal and geographic variations in those measures; we pay particular attention to California MSAs.

To the extent that extreme movements in MSA house price returns are few in number or geographically random, they would be of limited consequence to either private investors or policymakers. On the other hand, higher levels of ubiquity in return or jump return correlations raise concerns for mortgage or housing investors seeking to diversify risks associated with extreme house price movements. In a similar vein, other market players including MBS originators and investors would be similarly impacted by high correlations in returns or jump returns among their mortgage assets. Note further that jumps or jump correlations may be driven by economic or policy shocks at local or national levels. Jumps in house price returns should be of interest to policymakers especially in those cases where jumps can be traced to political events or policy perturbations.

Prior analyses have proposed alternative measures of jump test statistics (see, for example, Barndorff-Nielson and Shepard (2006), Lee and Mykland (2008), Jiang and Oomen (2008), and Jacod and Todorov (2009)). In a recent paper, Pukthuanthong-Le and Roll (2010) assess the various jump statistics in application to stock return indexes for 82 countries.[8] Unlike the other measures, Lee and Mykland works well with single observations (as opposed to a sample of several observations). This is important for our application because we have only quarterly data and hence the sample size is more limited than in the case of equities, where daily observations are available. While results vary across alternative jump statistics, results of of the above cited research suggest that jumps are largely idiosyncratic in international equity indexes. We are not aware of prior analyses of jumps in metropolitan house prices returns.

For the vast majority of sampled MSA housing markets, the most frequent quality-adjusted house price index available to investors is quarterly. Moreover, investor rebalancing of real estate portfolios tends to be of lower frequency relative to that of equities, and commonly is at a quarterly interval. Consequently, we view such frequency as appropriate to investor and policymaker market assessment and hence for the jump analysis.

[8] Earlier work on extreme returns and correlation of same focused on more ad-hoc approaches (see Longin and Solnik, 2001).

With that in mind, we apply the Lee and Mykland (2008), (hereafter LM), method in assessment of extreme movements in US metropolitan house price indexes. Like Barndorff-Nielson and Shephard (2006), Lee and Mykland's (2008) test is based on bipower variation. Bipower variation is used to proxy the instantaneous variance of the continuous non-jump component of prices.

To understand the test, consider the following notation:

t, subscript for quarter

$T_k$, the number of quarters in subperiod k

K, the total number of available subperiods

$R_{i,t,k}$, the return (log price relative) for MSA i quarter t in subperiod k

The Barndorff-Nielson and Shepard (2006) and Lee and Mykland (2008) bipower variation, $B_{i,k}$,is defined as follows:

$$B_{i,k} = \frac{1}{T_k - 1} \sum_{t=2}^{T_k} | R_{i,t,k} \| R_{i,t-1,k} |$$

LM suggest the computation of bipower variation using data preceding a particular return observation being tested for a jump. The test statistic is $L = R_{i,t+1,k} / \sqrt{B_{i,k}}$ . Under the null hypothesis of no jump at t+1, LM show that $L\sqrt{2/\pi}$ converges to a unit normal. In addition, if there is a jump at t+1, $L\sqrt{2/\pi}$ is equal to a unit normal plus the jump scaled by the standard deviation of the continuous portion of the process.

Jumps in housing returns, although frequent, do not occur as often as in equity returns (see Roll and Pukthuanthong-Le (2010)). In Figure 3, we describe the temporal incidence of big LM jumps in house price returns for US MSAs. For each quarter, we plot the percentage of LM statistics in excess of 2.0. That percentage is plotted from 1983:Q4 – 2010:Q1. Since the L statistic is asymptotically unit normal, we adopt a 10 percent criterion for each tail. In other words, we identify a non-normal (jump) quarter for each MSA when the absolute value of the LM statistic exceeds the 10 percent level for the unit normal (1.65).

Panel A of Figure 3 plots the quarterly incidence of big LM jumps for the full sample of 384 MSAs. Some evidence of jumps in house price returns is indicated for the overheated housing markets of the late 1980s with an incidence rate often in excess of 10 percent. Jumps fell back during the downturn of the early 1990s and were similarly damped from the mid-1990s through about 2003. In fact, results indicate a large number of quarters during the 1995 – 2003 period for which few if any US MSAs were characterized by statistical jumps in house prices returns.

As is evident, the 2000s bubble period was characterized by substantial jump incidence. Jumps were especially evident early in the boom during 2004-2005 as well as in 2008 in the wake of the bust in house prices. The latter set of jumps likely was associated with extreme declines in house price returns in a small percentage of metropolitan areas.

As in the above integration analysis, we assess jumps across inland and coastal California MSAs (Figure 3, panel B). In contrast to the US as a whole, analysis for within California suggests virtually no statistical jumps in house price returns prior to 2003. However, during the early stages of the boom period (2003 – 2004), return jumps suddenly became very prevalent with close to 70 percent having significant extreme returns. The jumps in returns were evidenced among both coastal and inland California cities; indeed, the plots reveal little difference in either the timing or incidence of house price jumps among MSAs in those areas. In marked contrast, substantially elevated incidence of significant extreme values (LM return jumps) was indicated during the bust 2007-2008 period only for inland California MSAs! Indeed, there is no evidence of jumps in returns during the latter period for coastal cities. The jumps evidenced for inland California cities during the bust period likely reflect the sharp house price declines that were common in those areas. Such outcomes were consistent with the implosion in housing market drivers. As suggested above, unlike coastal areas, inland cities were characterized by lack of (regulatory or natural) constraint on housing supply and were substantially overbuilt. Further, inland areas shared a common feature of substantial boom period subprime lending. As boom turned to bust, inland areas of California quickly and largely imploded. While the preceding indicates the marked incidence of house price return jumps during the 2000s housing boom and bust, they provide little insight as regards contemporaneous or lagged MSA correlations in those jumps, and returns in general. It is to those analyses that we now turn.

First, a word on methodology. Per above and following Pukthuanthong-Le and Roll (2010), we identify periods when the L statistic indicates a likely jump. After classifying each sample quarter for each MSA as jump or non-jump (jump indicated in those cases where the absolute value of the LM L statistics is greater than 2.0, given that L is unit normal), we compute contemporaneous and lagged correlations in LM jump statistics among pairs of MSAs where at least one MSA had a jump. If the companion MSA also had a jump in the same quarter (or in the lagged quarter) the product of their LM measures contributes to the contemporaneous (or lagged) correlation. Otherwise, the contribution for that month is zero. Note that we do not count the LM statistic for a given quarter unless it is significant; this is appropriate, otherwise the resulting correlation would simply measure the total return correlation. The result of our procedure is a pure measure of jump correlation for every pair of MSAs.

We find extensive evidence of strong correlations in returns and jumps. But jumps occur infrequently and have smaller correlations than returns. California exhibits particularly large return and jump correlations. In Table 3, we report summary information on MSA house price return and jump return correlations. Panel A reports summary statistics for MSA return correlations, which provide a basis of comparison to MSA jump correlations. Those results are stratified by level of T-statistic for cross-coefficient independence. For the full sample, correlation coefficients are computed for quarterly returns among all house price return pairs (total sample N = 73,536). The mean contemporaneous correlation among all MSAs return pairs is 0.20, with considerable cross coefficient standard deviation of 0.18. However, the T-statistic for the mean correlation, assuming cross-coefficient independence, is almost 300, indicating very significant average correlation among MSA returns. The table further indicates sizable numbers of individual MSA pairs with house price return correlations at high levels of statistical significance. The numbers of MSA pairs with return correlation T-statistics in excess of 2 and 3 are 33,460 and 18,126, respectively. Among those same sub-samples, mean correlations are 0.35 and 0.44, respectively.

Panel B of Table 3 reports summary statistics for the corresponding jump return correlations stratified by T-statistic. For the full sample, correlation coefficients are computed for identified jumps in quarterly house price returns among US MSAs. There are 49,742 pairs. The summary statistics are computed across all available coefficients. The mean contemporaneous MSA jump correlation across MSA jump return pairs is only about 0.05 but is significant with a T-statistic of about 53. The Table further indicates the existence of MSA house price jump return correlations at higher levels of statistical significance. The numbers of MSA pairs with jump return correlation T-statistics in excess of 2 and 3 are 8770 and 5405, respectively. Among these more significant sub-samples, mean correlations as expected are substantially higher (0.38 and 0.46, respectively.) And these samples are similarly characterized by significant MSA jump mean correlations, as indicated by T-statistics of 237 and 247, respectively.

We now turn to identify the geographical incidence of significant return and jump correlations in metropolitan housing returns. We find strong evidence for a high incidence of significant return and jump return correlations for California. In panel A of Table 4, contemporaneous and lead MSA house price index return correlations coefficients are computed for US census divisions. In that analysis, we break out California MSAs. Accordingly, the definition of census division 1 is now non-standard, as we remove California from that division. As is evident in the top left-hand panel, the incidence of MSA house price return correlations varies substantially across US census divisions. For each division, the number and proportion of significant correlations (using a T-stat of 5 or above) are reported. The mean correlation for each region is also given. The vast majority of census divisions, including divisions 1 – 8, report only limited contemporaneous correlations in MSA house price returns. Specifically, divisions 1 – 8 report a mean correlation coefficient in the range of 0.2 – 0.3 with not more than around 20 percent highly significant. California appears to be different from the rest of the U.S. in that 92 percent of the MSA paired returns are significantly contemporaneously correlated! Further, the mean correlation level for California MSAs is about .66!

As reported in the top right-hand panel of table 4, intertemporal (lead one quarter ahead) correlations are similarly damped in most census divisions. Among divisions 1 - 8, less than 10 percent of lead correlations are statistically significant. Further, mean lead correlation levels remain at or below .20. In marked contrast, MSAs in New England (division 9) and California are characterized by relatively high percentages of significant and elevated lead correlations. Again California is the outlier, as in excess of three-quarters of California MSAs recorded significant lead return correlations with a mean correlation level of about .57.

Panel B reports a similar assessment of contemporaneous and lead LM jump return correlations among MSAs stratified by census division. As shown in the bottom panels, California is conspicuously different from the rest of the U.S. For census divisions 1 – 8, significant contemporaneous jump correlations are small in number (less than 10 percent in any division) and mean correlations coefficients are in the range of only .02 – .03. In those same areas, lead jump correlations are limited to an incidence of 6 percent or less in any division with mean correlation coefficients (except for New England) of .04 or less. In marked contrast, jump return contemporaneous correlations are significant among California MSAs at an occurrence rate of 34 percent, and with much larger values, reaching .22, substantially in excess of levels discussed above for other regions. Moreover, the mean lead jump correlations are highest for California.

Another clear message results from the correlation analysis in US housing markets and when broken down into geographical cohorts. The incidence of significant return correlations far exceeds jump correlations. To illustrate, the percentage with significant t-statistics greater than 2 is in excess of 45 percent for return correlations compared to approximately 18 percent for jump return correlations (see Table 3). When we break out the analysis into geographical cohorts we find that the ratio of significant t-statistics far greater for return correlations with three exceptions, that occur in Divisions 3 through 5 for lead values (see Table 4). The results pertaining to the magnitude of correlations across return and jump returns are even more clear-cut. In all comparisons, we find that the return correlations far exceed their jump counterparts, usually by a ratio of 5 or more!

In addition, analyses of contemporaneous and lead jumps in house price returns again suggest that California is different. Also, levels of contemporaneous and lead return and jump correlations in California were well in excess of levels recorded in other census divisions. Given the anomalous behavior of California metropolitan housing markets thus documented we now turn to identify further insights as regards the temporal – spatial structure of house price return contagion in this state.

## IV. Contagion in Housing Market Returns

The above analyses suggest the outlier status of California MSAs in assessment of recent house price phenomena. Specifically, our analyses point to rising levels of integration as well as elevated return correlation and jump return correlation, both lead and contemporaneous, among California MSAs. However, the spatial dimensions of those relationships were not specified. Below we address that issue via parametric assessment of the spatial dynamics of housing returns among MSAs in northern and southern California.

We report some interesting findings for the metropolitan housing markets in California. In particular, spatial return spillovers are largely efficient across MSAs, especially in Southern California, coming from Los Angeles to surrounding areas. Results of a first set of analyses are contained in Table 5. There we test the simple hypothesis that house price returns among primary California coastal MSAs lead those of surrounding areas. That hypothesis is consistent with a mechanism whereby increases in house price returns (and related declines in affordability) in expensive, supply-constrained, coastal metropolitan areas lead to out-migration, related demand-side pressures, and subsequent increases in returns in more affordable inland suburbs. In our test of that hypothesis for southern California, for example, we estimate city-specific regressions whereby we regress returns for each inner- and outer-ring suburb of the larger LA area on contemporaneous and lead Los Angeles MSA house price returns. We undertake identical analyses for the Bay Area and central California using San Francisco and Santa Barbara as primary coastal cities. As shown in Table 5, we estimate those equations over the full timeframe of the metro-specific data sets. In each case, MSA returns are regressed on contemporaneous and 3 quarterly lags of primary coastal MSA returns.

Results of the analysis for LA region MSAs are contained in the top panel of Table 5. Those findings indicate a market efficiency in metropolitan spillover returns in that the most significant effects are contemporaneous. Overall, the regressions are characterized by high levels of explanatory power. In all of LA's surrounding cities, including Bakersfield, Fresno, Oxnard-Thousand Oaks, Riverside, San Diego, Santa Ana, and Santa Barbara, sizable and highly significant coefficients are estimated for contemporaneous Los Angeles house price returns. In Bakersfield and Fresno, located further from Los Angeles in California's great central valley, the contemporaneous coefficients on Los Angeles house price returns are about .60 and highly significant; further, a positive and significant coefficient of about .30 is estimated on the first quarterly lag of Los Angeles house price returns. In marked contrast, in closer-in areas, only the contemporaneous coefficient was statistically significant. Indeed, in those cities, the estimated coefficients on contemporaneous (quarterly) changes in Los Angeles house price returns were close to 1! These analyses indicate a high degree of contemporaneous correlation in house price returns among Los Angeles and its suburbs.

Results of the analysis diverge somewhat for San Francisco and environs where the level of market efficiency appears to be somewhat lower. In most areas of northern California, including Oakland, Sacramento, Salinas, San Jose, Santa Rosa, and Santa Cruz, both contemporaneous and 1-quarter lagged San Francisco house price returns play a sizable and significant role in determination of house price returns. In a few places, including both Oakland and Santa Cruz, contemporaneous as well as 1- and 2-quarter lagged San Francisco house prices returns significantly affect surrounding outcomes. San Francisco house price returns lead those of the outer-ring Central Valley boom town of Modesto by 1-quarter. In short, findings for Bay Area regional housing markets suggest a spatial term structure of contagion, whereas results for Los Angeles indicate a southern California region where metropolitan housing returns largely move in lock-step.

The above findings, however, may not be robust to periods of boom and bust in California housing markets. Indeed, it is plausible that the spatial or temporal path of house price contagion might accelerate during a boom or decelerate and even reverse during a bust. We test for such effects in Table 6. The regression equations estimated in Table 6 are identical to those in Table 5, except that each regression contains 4 additional terms. The additional variables comprise interactions between the primary (explanatory) city's return (contemporaneous and 3 quarterly lags) and a contemporaneous residual from a time trend fit of the log of an equal-weighted index of California house prices.

Findings contained in table 6 indicate that results of the California MSA house price contagion analysis are largely robust to the inclusion of the boom and bust interactive terms. In southern California, an exception is Bakersfield, where a sizable and significant coefficient is estimated on second quarterly lagged interaction term. In northern California, there exists little to report other than significant coefficients on contemporaneous interactive terms for Santa Rosa and Santa Cruz. Accordingly, an explicit accounting for boom and bust periods in California's housing markets has little effect on conclusions regarding the temporal path of house price contagion among California MSAs.[9]

## V. Conclusion

This paper applies data from 384 US MSAs to examine integration and contagion among metropolitan housing markets. The paper first examines the level and change in housing market integration as reflected in the response of MSA house price returns to a national multi-factor model. It then investigates the incidence of large house price return and jump return correlations for the MSAs. Finally, as a result of the earlier integration and contagion analysis, it isolates California and further examines contagion characteristics from leading coastal cities to their inland neighbors.

Research findings reveal a highly integrated set of US metropolitan housing markets. Furthermore, the susceptibility of MSA housing markets to national economic and policy shocks trended up over time and was especially evident in the decade of the 2000s. Also, high levels and elevated trends in housing market integration limit the efficacy of strategies to diversify MSA-specific risk on the part of mortgage and housing investors.

[9] We undertook yet another robustness check whereby we created an interaction between the explanatory's city's return (including four lags) and a contemporaneous residual from a time trend fit of the log of an equal-weighted California MSA (N=28) FHFA house price index. That interaction term was substituted for the primary coastal city boom and bust interaction term estimated in Table 6. Results here differed little from those reported in table 6, as the house price index for the state as a whole differed little from those for the primary coastal California cities.

California emerges as somewhat of an outlier, in terms of elevated trends in integration, jumps in house price returns, and MSA contemporaneous and lagged return and jump return correlations. In addition, high levels of short-term contagion appear endemic to major California markets, especially in Los Angeles. Inland California MSAs appear to behave as one and exhibit a high degree of market efficiency in the response to return movements in the large coastal metropolitan areas.

**Figure 1: US and California House Price Indices**

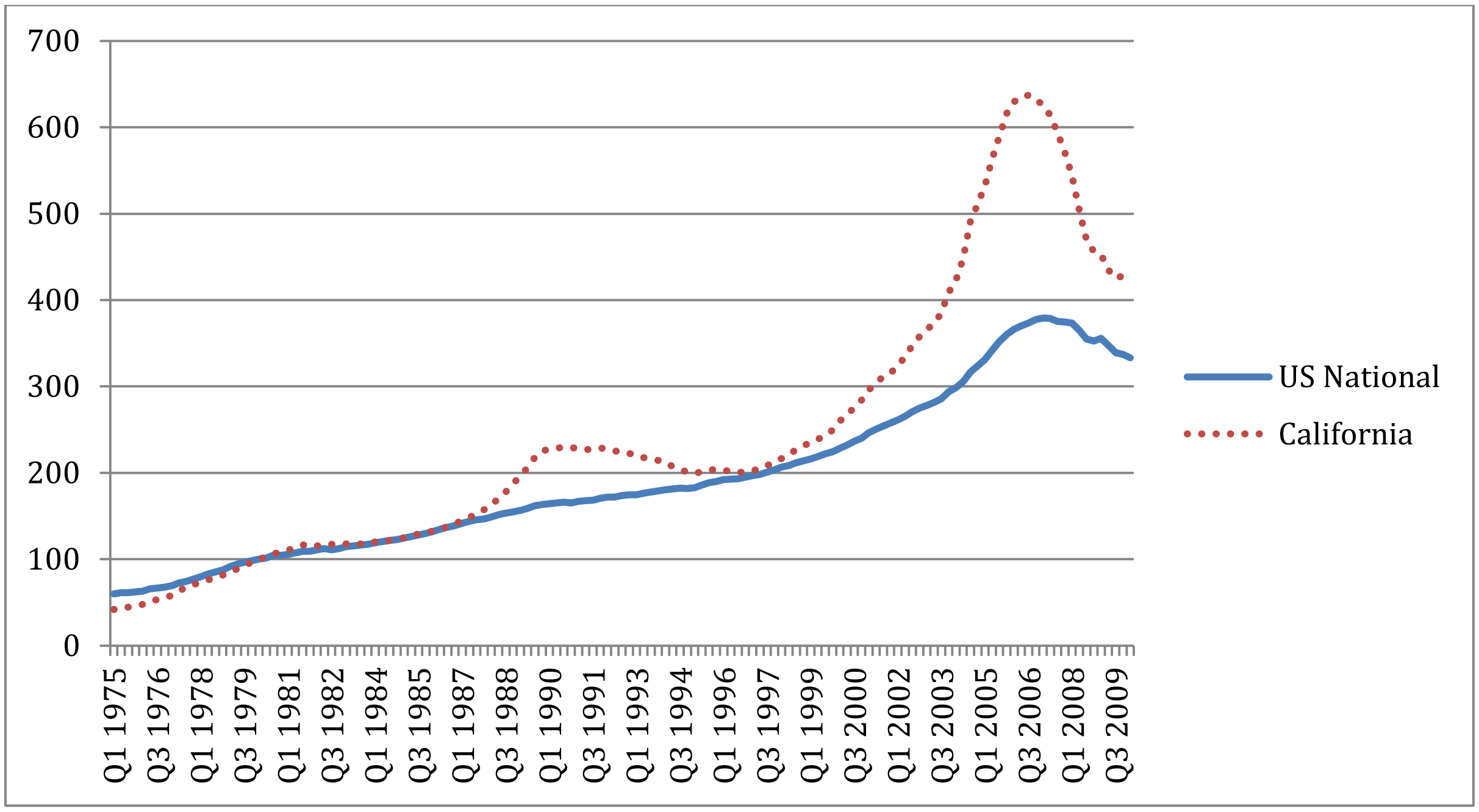

Notes: The chart depicts the time series of US national and California index levels (1975: Q1 - 2010:Q1) based on repeat sales house price indexes from the Federal Housing Finance Agency (FHFA). The prices are normalized to 100 in 1980:Q1.

**Figure 2: Housing Return Integration Trends**
**Panel A: Average R-squares for US MSAs and California MSAs**

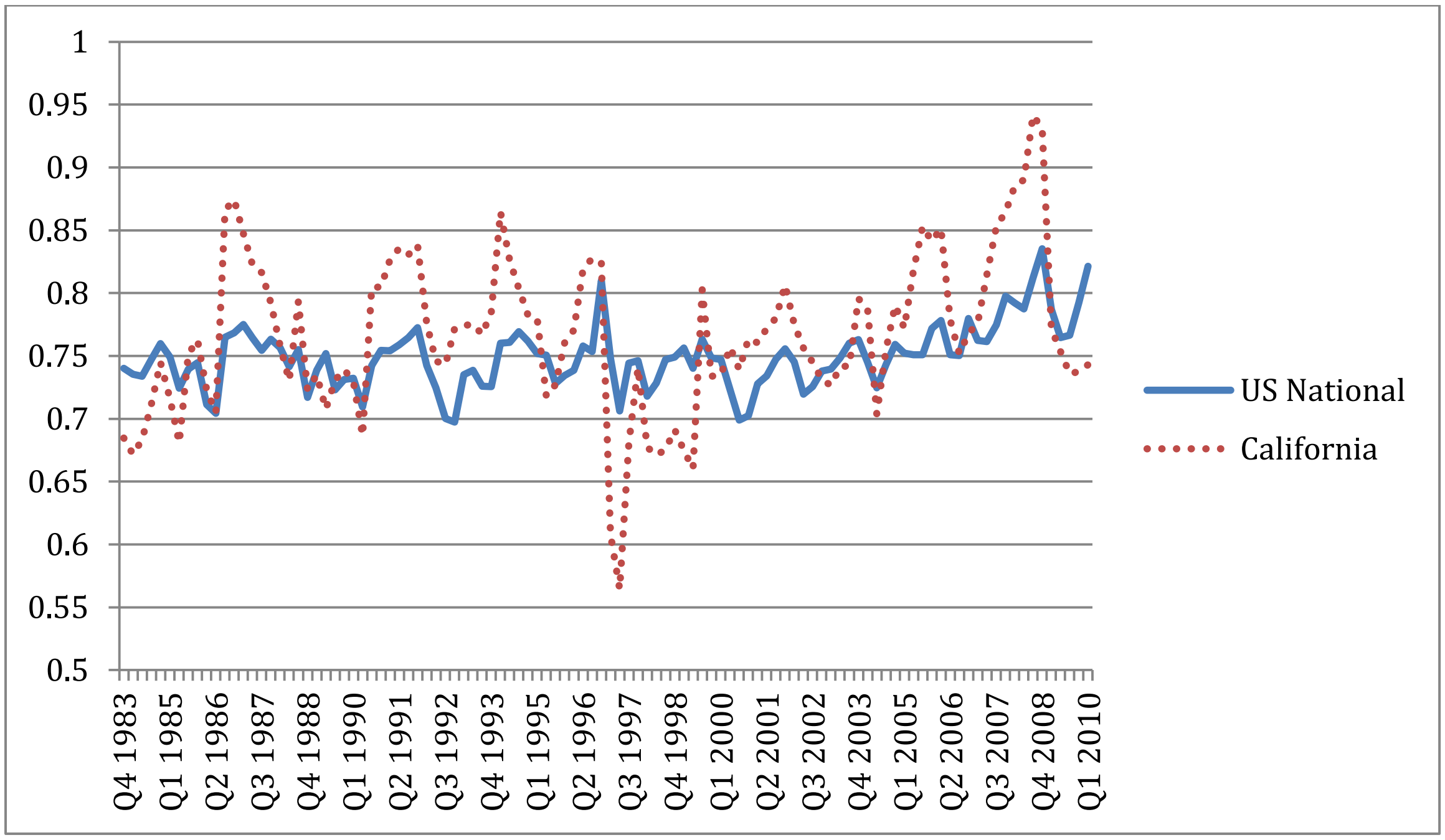

**Panel B: Average R-squares for US MSA Time Cohorts**

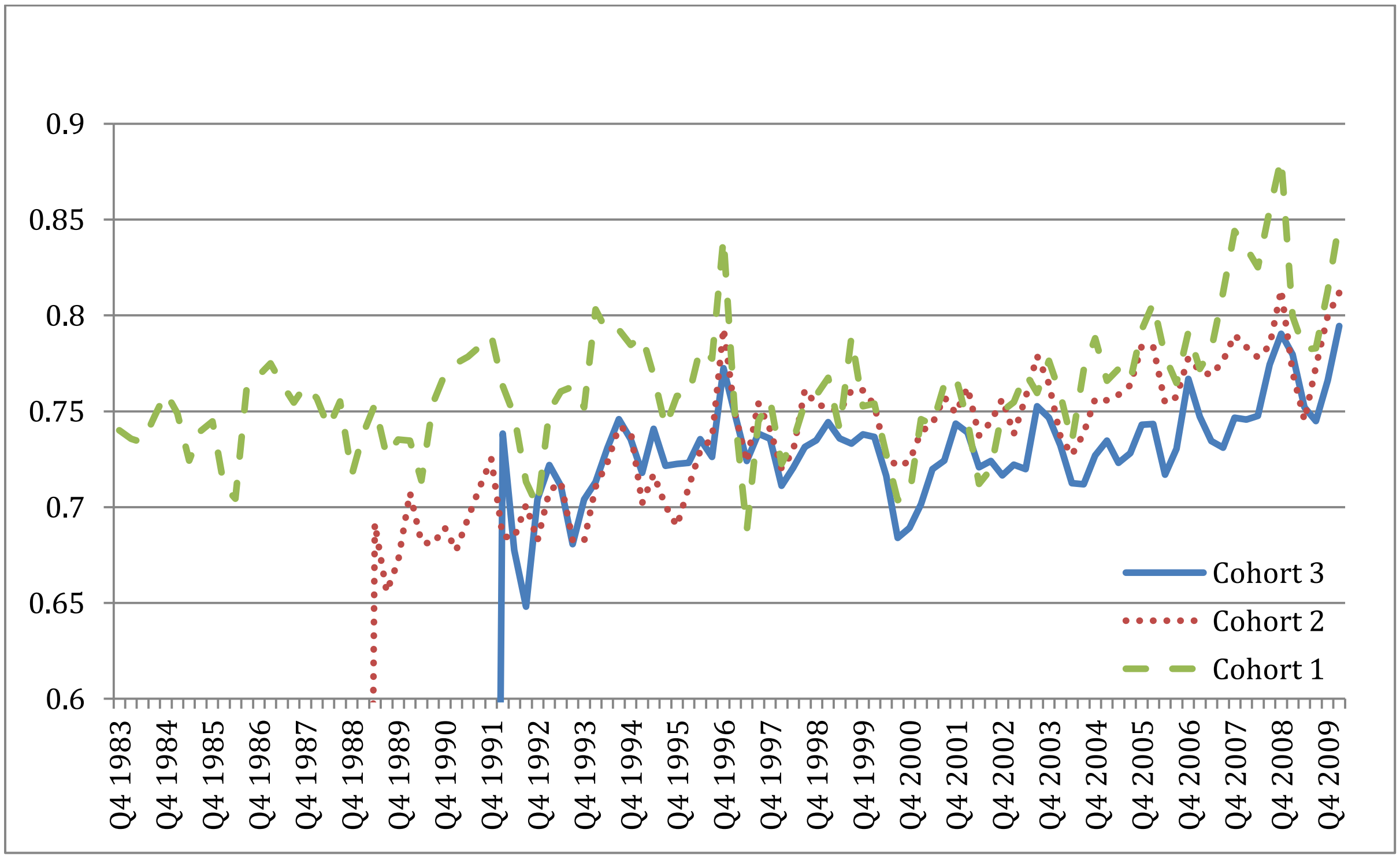

**Panel C: Average R-squares for California Inland and Coastal MSAs**

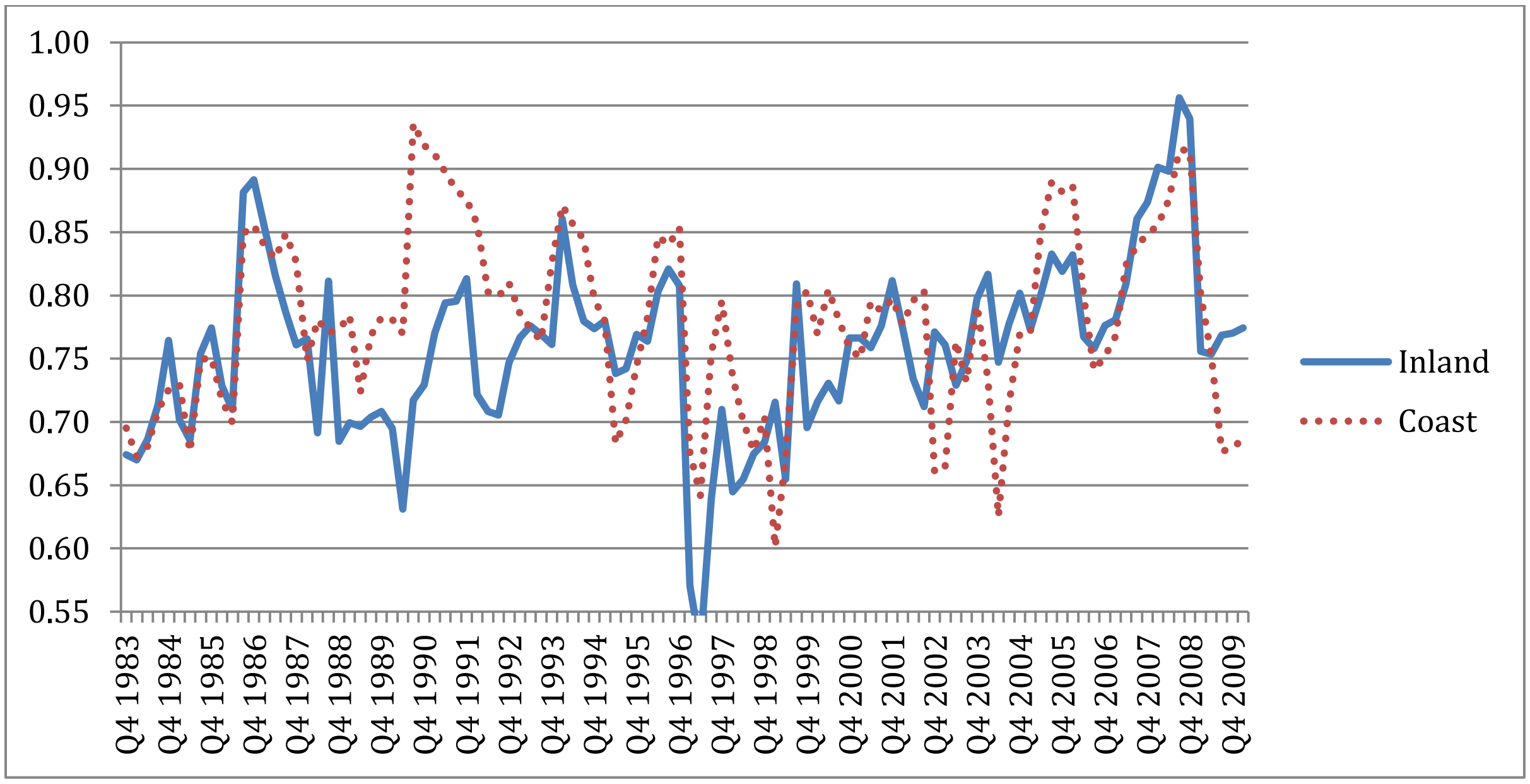

Notes: The level of integration is measured by the R-squares from the multi-factor housing returns model fitted for the full sample of MSAs using a 20-quarter moving window. See Appendix Table 1 for details on the factors utilized in model estimation. Average levels of integration are presented for 1983:Q4 – 2010:Q1 for 384 US MSAs and for 28 California MSAs. Average levels of integration are presented for time cohorts based on when the MSA entered the database and had sufficient time series to execute the moving window regression. The cohorts begin at 1983: Q4 (cohort 1), 1989:Q2 (cohort 2) and 1992:Q1 (cohort 3). Average levels of integration are also presented for California Interior MSAs and California Coast MSAs. California Coastal MSAs include Los Angeles, Oakland, Oxnard, San Diego, San Francisco, San Jose, San Luis Obispo, Santa Ana, Santa Barbara and Santa Cruz with the remainder of the 28 MSAs categorized as California Inland MSAs.

**Figure 3: US and California LM Jump Statistics**
**Panel A: Big LM House Price Return Jumps Proportion [% |LM| > 2] for US MSAs by Quarter**

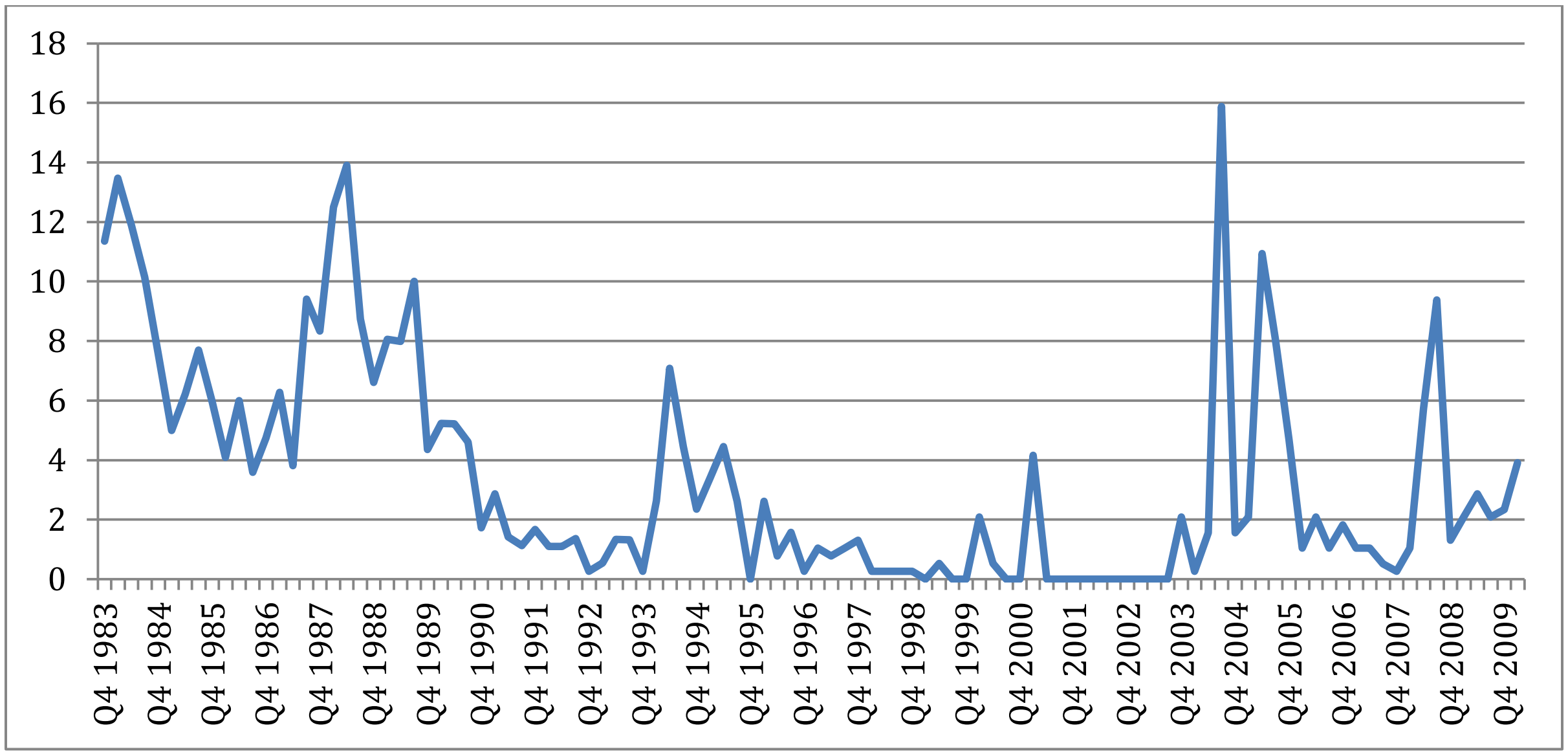

**Panel B: Big LM House Price Return Jumps Proportion [% |LM| > 2] for Coastal and Inland California MSAs by Quarter**

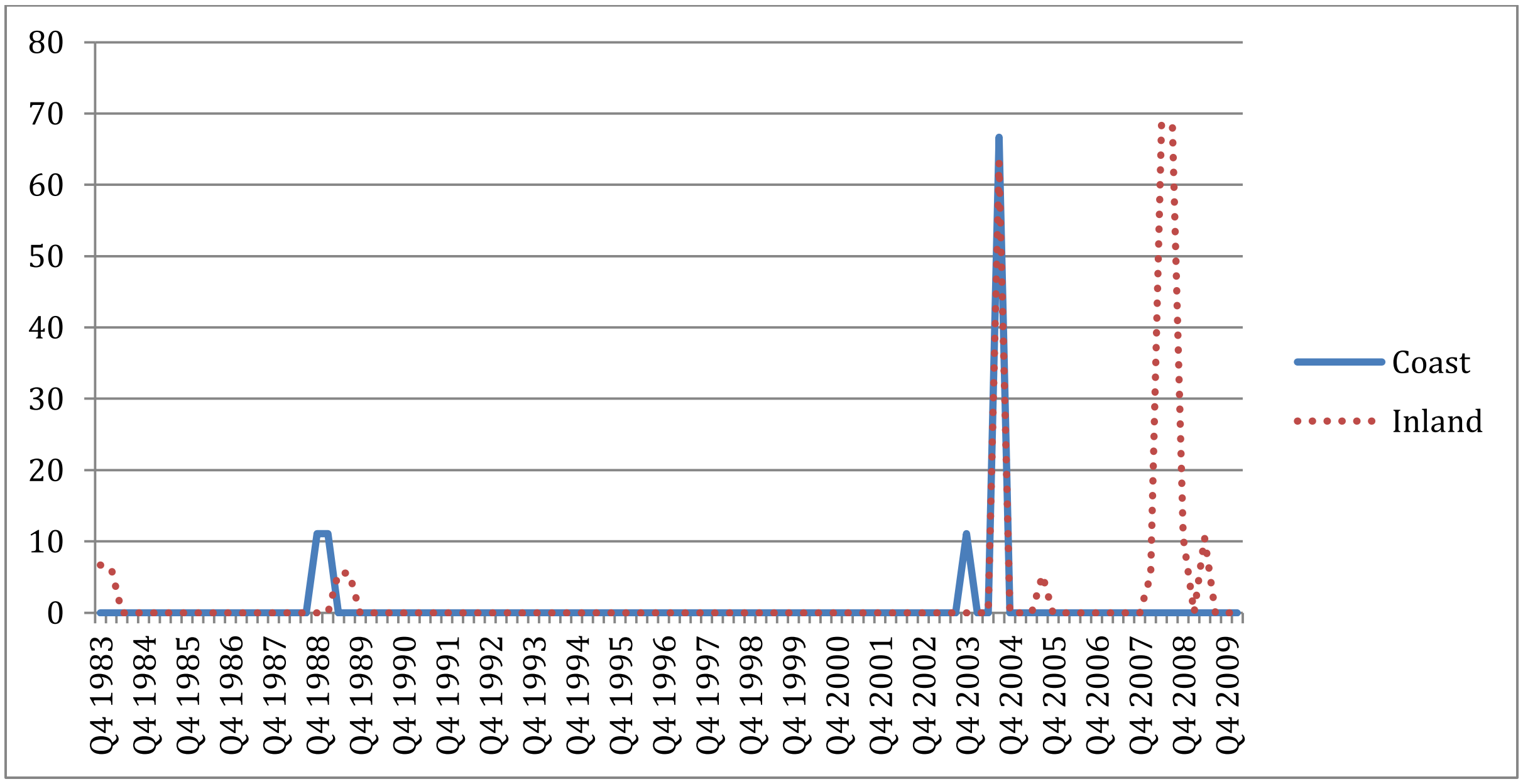

Notes: The Lee and Mykland (2008) (LM) jump measure is computed from quarterly observations for each of the 384 MSAs. Plots are given for the US National, and for inland and coast California MSAs. The plots are from 1983:Q4 and show the percentage of LM statistic that exceed 2.0. The percentage classified as a jump quarter is when the absolute value of the LM statistic exceeds the 10% level for a unit normal (1.65).

## Table 1
## Summary Integration Measures for All MSAs

| | Mean | Sigma | Final R-Square | Change in R-Square | R-Square Trend T-stat |
|---|---|---|---|---|---|
| Mean | 0.988 | 2.450 | 0.822 | 0.093 | 1.222 |
| Std Dev | 0.259 | 0.890 | 0.118 | 0.187 | 2.879 |
| Min/Quintile 1 | 0.430 | 0.980 | 0.349 | -0.616 | -7.246 |
| Quintile 2 | 0.784 | 1.744 | 0.738 | -0.046 | -1.167 |
| Quintile 3 | 0.890 | 2.144 | 0.817 | 0.053 | 0.501 |
| Quintile 4 | 0.998 | 2.545 | 0.864 | 0.120 | 2.035 |
| Quintile 5 | 1.185 | 2.958 | 0.930 | 0.236 | 3.436 |
| Max | 1.892 | 9.258 | 0.993 | 0.695 | 10.469 |

Summary details for 5 integration characteristics (Mean, Sigma, Final R-square, Change in R-square, and R-Square Trend T-stat) are presented for the 384 MSAs. Mean is the average quarterly house price return. We compute house price returns for each MSA in our sample as the log quarterly difference in its FHFA repeat home sales price index. Sigma is the standard deviation of returns. We use R-Squares as the measure of integration and these are applied to obtain R-square trend t-statistics. R-squares are obtained from fitting MSA returns to the factors described in Appendix Table 1. The time trend t-statistics are estimated by regressing the R-squares for each MSA on a simple linear time trend for all available quarters of data. The final R-squares pertain to 2010:Q1 for all 384 US MSAs. The change in R-squares refers to the difference between estimates for 2010:Q1 and 1983:Q4 for each MSA. Summary details report the time-series cross-sectional summary statistics (mean, standard deviation, minimum/quintile 1, quintile 2, quintile 3, quintile 4, quintile 5 and maximum) of the characteristics. The minimum values of each quintile are presented.

## Table 2
## Summary Integration Measures for California MSAs

| MSA | Mean | US Rank Mean | CA Rank Mean | Sigma | US Rank Sigma | CA Rank Sigma | Final R-Square | Change in R-Square | Trend t-stat | US Rank Trend t-stat | CA Rank Trend t-stat |
|---|---|---|---|---|---|---|---|---|---|---|---|
| Bakersfield | 0.864 | 136 | 4 | 3.197 | 330 | 16 | 0.898 | 0.166 | 4.228 | 335 | 26 |
| Chico | 1.066 | 273 | 11 | 3.077 | 321 | 13 | 0.832 | 0.169 | -0.844 | 87 | 7 |
| El Centro | 0.607 | 11 | 1 | 4.240 | 370 | 27 | 0.912 | 0.114 | 2.365 | 258 | 18 |
| Fresno | 1.075 | 276 | 12 | 3.198 | 331 | 17 | 0.833 | -0.004 | 2.174 | 241 | 16 |
| Hanford | 0.909 | 172 | 8 | 3.098 | 324 | 15 | 0.619 | 0.226 | 4.120 | 331 | 25 |
| Los Angeles | 1.736 | 380 | 26 | 2.839 | 286 | 5 | 0.558 | -0.339 | 2.172 | 239 | 15 |
| Madera | 0.879 | 146 | 6 | 3.548 | 351 | 23 | 0.826 | -0.121 | 8.208 | 380 | 28 |
| Merced | 0.790 | 84 | 2 | 4.674 | 376 | 28 | 0.889 | 0.111 | 2.937 | 282 | 20 |
| Modesto | 1.005 | 236 | 9 | 4.006 | 364 | 26 | 0.820 | 0.168 | 2.994 | 286 | 22 |
| Napa | 1.424 | 358 | 18 | 2.989 | 312 | 9 | 0.838 | 0.158 | 3.463 | 310 | 24 |
| Oakland | 1.699 | 378 | 25 | 2.638 | 250 | 2 | 0.577 | -0.115 | 0.744 | 167 | 12 |
| Oxnard | 1.635 | 374 | 23 | 2.991 | 313 | 10 | 0.768 | 0.099 | 2.353 | 255 | 17 |
| Redding | 0.879 | 148 | 7 | 3.063 | 319 | 12 | 0.947 | 0.388 | -1.395 | 68 | 5 |
| Riverside | 1.296 | 332 | 14 | 3.438 | 343 | 21 | 0.713 | 0.177 | -1.260 | 74 | 6 |
| Sacramento | 1.354 | 345 | 17 | 2.894 | 299 | 8 | 0.649 | -0.176 | 2.980 | 284 | 21 |
| San Diego | 1.541 | 369 | 16 | 3.013 | 314 | 24 | 0.868 | -0.114 | 0.253 | 141 | 14 |
| San Francisco | 1.892 | 384 | 20 | 2.540 | 229 | 11 | 0.638 | -0.143 | -2.069 | 50 | 10 |
| San Jose | 1.877 | 383 | 28 | 2.789 | 274 | 1 | 0.759 | 0.123 | -2.379 | 42 | 3 |
| San Luis Obispo | 1.303 | 334 | 27 | 3.326 | 337 | 4 | 0.637 | -0.134 | -0.826 | 88 | 1 |
| Santa Ana | 1.674 | 376 | 15 | 2.718 | 265 | 19 | 0.626 | 0.051 | -1.713 | 58 | 8 |
| Santa Barbara | 1.470 | 364 | 24 | 2.879 | 295 | 3 | 0.779 | 0.090 | 0.256 | 142 | 4 |
| Santa Cruz | 1.599 | 373 | 19 | 3.093 | 323 | 7 | 0.657 | 0.065 | 2.422 | 262 | 11 |
| Santa Rosa | 1.590 | 371 | 22 | 2.855 | 291 | 14 | 0.678 | 0.251 | 0.990 | 182 | 19 |
| Stockton | 1.050 | 266 | 21 | 3.696 | 359 | 6 | 0.669 | 0.238 | -0.537 | 108 | 13 |
| Vallejo | 1.133 | 293 | 10 | 3.419 | 342 | 25 | 0.796 | 0.074 | -2.146 | 47 | 9 |
| Visalia | 0.872 | 142 | 13 | 3.244 | 334 | 20 | 0.828 | -0.013 | 3.380 | 303 | 2 |
| Yuba City | 0.833 | 113 | 5 | 3.448 | 345 | 18 | 0.579 | 0.030 | 5.775 | 361 | 23 |
| | ***Mean*** | | | ***Sigma*** | | | ***Final R-Square*** | ***Change in R-Square*** | ***Trend t-stat*** | | |
| Mean | 1.264 | | | 3.231 | | | 0.741 | 0.057 | 1.447 | | |
| Std Dev | 0.368 | | | 0.485 | | | 0.118 | 0.158 | 2.590 | | |
| Min | 0.607 | | | 2.540 | | | 0.558 | -0.339 | -2.379 | | |
| Max | 1.892 | | | 4.674 | | | 0.947 | 0.388 | 8.208 | | |

Notes: Details for 3 integration characteristics (Mean, Sigma and R-Square Trend t-stat) are presented for all 28 California MSAs. Mean is the average quarterly house price return. We compute house price returns for each MSA in our sample as the log quarterly difference in its FHFA repeat home sales price index. Sigma is the standard deviation of returns. R-Squares are the estimates of integration and are used to obtain R-Square trend t-statistics. R-squares are obtained from fitting MSA returns to the factor model described in Appendix Table 1. The time trend t-statistics are estimated by regressing the R-squares for each MSA on a simple linear time trend for all available quarters of data. The final R-Squares pertain to 2010:Q1 for all 28 California MSAs. The change in R-Squares refers to the difference between estimates for 2010:Q1 and 1983:Q4 for each MSA. Each characteristic is ranked from lowest to highest in comparison both to all 384 US MSAs and all 28 California MSAs. The last four rows provide the time-series cross-sectional summary statistics (mean, standard deviation, minimum and maximum) of the characteristics with reference to all CA MSAs.

**Table 3—MSA House Price Return and Jump Correlations**

Panel A: Return Correlations

| Full sample | | | | | |
|---|---|---|---|---|---|
| **N** | **Mean** | **Sigma** | **T-Stat** | **Maximum** | **Minimum** |
| 73536 | 0.201 | 0.182 | 299.735 | 0.946 | -0.639 |
| **Sample of correlations with T-statistic > 2** | | | | | |
| **N** | **Mean** | **Sigma** | **T-Stat** | **Maximum** | **Minimum** |
| 33460 | 0.354 | 0.125 | 517.703 | 0.946 | 0.173 |
| **Sample of correlations with T-statistic > 3** | | | | | |
| **N** | **Mean** | **Sigma** | **T-Stat** | **Maximum** | **Minimum** |
| 18126 | 0.435 | 0.116 | 505.922 | 0.946 | 0.258 |

**Table 3—MSA House Price Return and Jump Correlations**

Panel B: Jump Correlations

| Full sample | | | | | |
|---|---|---|---|---|---|
| **N** | **Mean** | **Sigma** | **T-Stat** | **Maximum** | **Minimum** |
| 49742 | 0.047 | 0.194 | 53.528 | 1.000 | -0.924 |
| **Sample of jump correlations with T-statistic > 2** | | | | | |
| **N** | **Mean** | **Sigma** | **T-Stat** | **Maximum** | **Minimum** |
| 8770 | 0.375 | 0.148 | 236.908 | 1.000 | 0.173 |
| **Sample of jump correlations with T-statistic > 3** | | | | | |
| **N** | **Mean** | **Sigma** | **T-Stat** | **Maximum** | **Minimum** |
| 5405 | 0.455 | 0.135 | 247.201 | 1.000 | 0.259 |

Notes: Notes: Panel A shows the house price return correlations. Correlation coefficients are computed from quarterly returns for all pairs of 384 MSAs (total sample N = 73536). Sigma is the cross-coefficient standard deviation. T is the T-statistic that tests for cross-coefficient independence. Panel B shows the jump correlations. Correlation coefficients are computed from quarterly returns for Lee and Mykland's (2008) (LM) jump measure. Sigma is the cross-coefficient standard deviation. T is the T-statistic that tests for cross-coefficient independence.

## Table 4
## Contemporaneous and Lagged MSA House Price Return and Jump Correlations by Geographical Cohort

Panel A: Return Correlations

| | Contemporaneous correlation | | | | Lead correlation | | | |
|---|---|---|---|---|---|---|---|---|
| | N | Number Significant | Percentage Significant | Mean Correlation | N | Number Significant | Percentage Significant | Mean Correlation |
| Division 1 | 190 | 37 | 19.474 | 0.304 | 400 | 34 | 8.500 | 0.182 |
| Division 2 | 595 | 83 | 13.950 | 0.314 | 1225 | 88 | 7.184 | 0.222 |
| Division 3 | 496 | 14 | 2.823 | 0.211 | 1024 | 6 | 0.586 | 0.100 |
| Division 4 | 903 | 29 | 3.212 | 0.180 | 1849 | 19 | 1.028 | 0.091 |
| Division 5 | 1953 | 129 | 6.605 | 0.268 | 3969 | 49 | 1.235 | 0.148 |
| Division 6 | 2628 | 237 | 9.018 | 0.251 | 5329 | 295 | 5.536 | 0.171 |
| Division 7 | 703 | 62 | 8.819 | 0.237 | 1444 | 62 | 4.294 | 0.154 |
| Division 8 | 561 | 100 | 17.825 | 0.317 | 1156 | 104 | 8.997 | 0.213 |
| Division 9 | 153 | 114 | 74.510 | 0.629 | 324 | 193 | 59.568 | 0.501 |
| CA | 378 | 349 | 92.328 | 0.656 | 784 | 596 | 76.020 | 0.565 |

## Table 4
## Contemporaneous and Lagged MSA House Price Return and Jump Correlations by Geographical Cohort

### Panel B: Jump (LM) Correlations

| | Contemporaneous correlation | | | | Lead correlation | | | |
|---|---|---|---|---|---|---|---|---|
| | N | Number Significant | Percentage Significant | Mean Correlation | N | Number Significant | Percentage Significant | Mean Correlation |
| Division 1 | 190 | 9 | 4.737 | 0.028 | 321 | 20 | 6.231 | -0.029 |
| Division 2 | 595 | 19 | 3.193 | 0.021 | 791 | 31 | 3.919 | 0.035 |
| Division 3 | 496 | 5 | 1.008 | 0.006 | 552 | 23 | 4.167 | 0.035 |
| Division 4 | 903 | 26 | 2.879 | 0.017 | 1124 | 60 | 5.338 | 0.025 |
| Division 5 | 1953 | 60 | 3.072 | 0.018 | 2479 | 111 | 4.478 | 0.037 |
| Division 6 | 2628 | 67 | 2.549 | 0.017 | 3772 | 84 | 2.227 | 0.034 |
| Division 7 | 703 | 33 | 4.694 | 0.033 | 1068 | 17 | 1.592 | 0.012 |
| Division 8 | 561 | 15 | 2.674 | 0.016 | 770 | 36 | 4.675 | 0.041 |
| Division 9 | 153 | 13 | 8.497 | 0.047 | 252 | 13 | 5.159 | 0.095 |
| CA | 378 | 130 | 34.392 | 0.224 | 705 | 49 | 6.950 | 0.116 |

Notes: Panel A presents the return correlations including both contemporaneous and lead (one quarter ahead) correlations. Correlation coefficients are computed from quarterly returns for each geographical division where N is the sample size. The number and proportion of significant correlations with a t-statistic greater than 5 are reported. The mean correlation is also given. Panel B presents the jump correlations including both contemporaneous and lead (one quarter ahead) correlations. Correlation coefficients are computed from quarterly returns for Lee and Mykland's (2008) (LM) jump measure for each geographical division where N is the sample size. The number and proportion of significant correlations with a t-statistic greater than 5 are reported. The mean correlation is also given. The geographical divisions are based on the 9 US census divisions. However the definition of division 1 is not standard, in that we remove California from census division 1 and report it separately in a cohort by itself (CA). The states in the 9 census divisions are: Division 1 (AK HI OR WA), Division 2 (AZ CO ID MT NM NV UT WY), Division 3 (IA KS MN MO ND NE SD), Division 4 (AR LA OK TX), Division 5 (IL IN MI OH WI), Division 6 (AL KY MS TN), Division 7 (DC DE FL GA MD NC SC VA WV), Division 8 (NJ NY PA) and Division 9 (CT MA ME NH RI VT).

## Table 5—Housing Return Contagion Regressions for California MSAs
### Panel A: Explanatory MSA - Los Angeles

| | N | Constant | Lag0 | Lag1 | Lag2 | Lag3 | R-Squares |
|---|---|---|---|---|---|---|---|
| Bakersfield | 130 | -0.460 | 0.600 | 0.350 | -0.210 | 0.120 | 0.516 |
| | | (-2.010) | (4.360) | (2.050) | (-1.220) | (0.880) | |
| | | | | | | | |
| | | | | | | | |
| Fresno | 131 | -0.280 | 0.600 | 0.290 | -0.230 | 0.190 | 0.509 |
| | | (-1.200) | (4.320) | (1.730) | (-1.340) | (1.400) | |
| Oxnard | 135 | 0.100 | 1.070 | 0.070 | -0.120 | -0.090 | 0.867 |
| | | (0.850) | (16.420) | (0.890) | (-1.430) | (-1.300) | |
| Riverside | 135 | -0.540 | 0.950 | 0.110 | 0.100 | -0.060 | 0.788 |
| | | (-3.300) | (10.100) | (0.970) | (0.810) | (-0.640) | |
| San Diego | 136 | 0.190 | 0.900 | -0.190 | 0.030 | 0.060 | 0.564 |
| | | (0.940) | (7.610) | (-1.270) | (0.200) | (0.510) | |
| Santa Ana | 136 | 0.100 | 0.900 | 0.020 | 0.010 | -0.010 | 0.895 |
| | | (1.040) | (17.170) | (0.250) | (0.140) | (-0.270) | |
| Santa Barbara | 129 | 0.220 | 0.890 | -0.080 | 0.070 | -0.050 | 0.649 |
| | | (1.240) | (8.390) | (-0.590) | (0.550) | (-0.450) | |

## Panel B: Explanatory MSA - San Francisco

| | N | Constant | Lag0 | Lag1 | Lag2 | Lag3 | R-Squares |
|---|---|---|---|---|---|---|---|
| Merced | 117 | -1.130 | 0.340 | 0.670 | 0.190 | -0.050 | 0.285 |
| | | (-2.430) | (1.190) | (2.100) | (0.690) | (-0.210) | |
| Modesto | 131 | -0.950 | -0.020 | 0.650 | 0.410 | 0.050 | 0.380 |
| | | (-2.640) | (-0.130) | (3.480) | (2.160) | (0.290) | |
| Napa | 125 | -0.100 | 0.370 | 0.130 | 0.400 | 0.010 | 0.439 |
| | | (-0.380) | (2.820) | (0.910) | (2.760) | (0.050) | |
| Oakland | 135 | -0.290 | 0.670 | 0.140 | 0.110 | 0.120 | 0.821 |
| | | (-2.280) | (11.330) | (2.170) | (1.710) | (2.090) | |
| Sacramento | 134 | -0.260 | 0.400 | 0.290 | -0.070 | 0.260 | 0.447 |
| | | (-1.050) | (3.500) | (2.310) | (-0.590) | (2.240) | |
| Salinas | 129 | -0.440 | 0.610 | 0.480 | -0.260 | 0.220 | 0.427 |
| | | (-1.430) | (3.880) | (2.880) | (-1.530) | (1.480) | |
| San Jose | 135 | -0.170 | 0.710 | 0.310 | 0.030 | 0.040 | 0.831 |
| | | (-1.350) | (11.750) | (4.620) | (0.380) | (0.730) | |
| Santa Cruz | 127 | -0.030 | 0.360 | 0.380 | 0.300 | -0.070 | 0.473 |
| | | (-0.130) | (2.780) | (2.610) | (2.060) | (-0.520) | |
| Santa Rosa | 133 | -0.230 | 0.440 | 0.410 | 0.070 | 0.090 | 0.623 |
| | | (-1.160) | (4.710) | (3.930) | (0.680) | (0.940) | |
| Stockton | 131 | -0.900 | 0.590 | 0.250 | 0.100 | 0.180 | 0.423 |
| | | (-2.840) | (3.790) | (1.510) | (0.570) | (1.190) | |
| Vallejo | 127 | -0.630 | 0.550 | 0.050 | 0.520 | -0.070 | 0.465 |
| | | (-2.240) | (3.790) | (0.310) | (3.240) | (-0.490) | |

**Panel C: Explanatory MSA - Santa Barbara**

| | | | | | | | |
|---|---|---|---|---|---|---|---|
| Oxnard | 126 | 0.040 | 0.550 | 0.270 | 0.130 | -0.020 | 0.673 |
| | | (0.250) | (7.400) | (3.700) | (1.750) | (-0.250) | |
| San Luis Obispo | 126 | 0.180 | 0.230 | 0.130 | 0.190 | 0.260 | 0.379 |
| | | (0.680) | (2.060) | (1.210) | (1.750) | (2.280) | |

Notes: Regression results for a selection of California MSAs on contemporaneous and lagged returns (3 lags) of large coastal California MSAs. The three large coastal leading cities used in the regressions are Los Angeles (Panel A), San Francesco (Panel B) and Santa Barbara (Panel C). N is the number of quarters in each regression. Regression coefficients and t-statistics in parentheses are given. R-squares of each regression are also reported. Results for some MSAs required a Cochrane-Orcutt adjustment for error term serial correlation. Durbin-Watson statistics for all presented MSA regressions allow us to reject the null hypothesis of first order serial correlation.

## Table 6— Housing Return Contagion Regressions Across Booms and Busts for California MSAs

### Panel A: Explanatory MSA - Los Angeles

| | N | Constant | Lag0 | Lag1 | Lag2 | Lag3 | Lag0 | Lag1 | Lag2 | Lag3 | R-Square |
|---|---|---|---|---|---|---|---|---|---|---|---|
| Bakersfield | 130 | -0.35 | 0.60 | 0.26 | -0.14 | 0.10 | 0.08 | 2.27 | -2.67 | 1.03 | 0.523 |
| | | (-1.34) | (3.87) | (1.49) | (-.81) | (0.68) | (0.07) | (1.64) | (-2.13) | (1.13) | |
| Fresno | 121 | 0.07 | 0.58 | -0.19 | 0.05 | 0.29 | 0.49 | -0.98 | -0.41 | 0.030 | 0.294 |
| | | (0.22) | (3.09) | (-0.93) | (0.22) | (1.58) | (0.36) | (-0.58) | (-0.25) | (0.03) | |
| Oxnard | 131 | -0.37 | 0.62 | 0.30 | -0.25 | 0.21 | -1.16 | -0.25 | 1.82 | -1.31 | 0.518 |
| | | (-1.47) | (4.06) | (1.76) | (-1.43) | (1.47) | (-1.03) | (-0.19) | (1.51) | (-1.43) | |
| Riverside | 135 | 0.14 | 1.01 | 0.10 | -0.14 | -0.050 | 0.26 | 0.010 | 0.13 | -0.96 | 0.882 |
| | | (1.210) | (15.55) | (1.32) | (-1.78) | (-0.80) | (0.67) | (0.02) | (0.25) | (-2.34) | |
| San Diego | 135 | -0.50 | 0.96 | 0.110 | 0.10 | -0.070 | -0.10 | 0.94 | -0.91 | 0.13 | 0.786 |
| | | (-2.80) | (9.44) | (0.91) | (0.83) | (-0.68) | (-0.16) | (1.18) | (-1.09) | (0.20) | |
| Santa Ana | 136 | 0.25 | 0.82 | -0.15 | -0.02 | 0.13 | 1.49 | -2.28 | 1.64 | -1.46 | 0.594 |
| | | (1.19) | (7.04) | (-1.03) | (-0.11) | (1.11) | (2.24) | (-2.46) | (1.83) | (-2.28) | |
| Santa Barbara | 136 | 0.13 | 0.87 | 0.030 | 0.00 | 0.00 | 0.11 | 0.34 | -0.31 | -0.51 | 0.904 |
| | | (1.40) | (16.83) | (0.50) | (-0.01) | (0.020) | (0.38) | (0.84) | (-0.77) | (-1.81) | |

## Panel B: Explanatory MSA - San Francisco

| | N | Constant | Lag0 | Lag1 | Lag2 | Lag3 | Lag0 | Lag1 | Lag2 | Lag3 | R-Square |
|---|---|---|---|---|---|---|---|---|---|---|---|
| Merced | 117 | -0.92 | 0.36 | 0.72 | -0.04 | 0.10 | 0.88 | 0.81 | 2.46 | -2.35 | 0.292 |
| | | (-1.88) | (1.18) | (1.90) | (-0.10) | (0.34) | (0.41) | (0.31) | (0.96) | (-1.19) | |
| Modesto | 131 | -0.84 | -0.01 | 0.67 | 0.39 | 0.06 | 0.15 | 1.51 | 0.51 | -1.33 | 0.387 |
| | | (-2.30) | (-0.08) | (3.31) | (1.85) | (0.33) | (0.13) | (1.27) | (0.39) | (-1.17) | |
| Napa | 125 | -0.10 | 0.55 | -0.18 | 0.46 | 0.09 | -2.05 | 3.18 | -0.33 | -1.18 | 0.458 |
| | | (-0.39) | (3.61) | (-0.92) | (2.29) | (0.56) | (-2.00) | (2.46) | (-0.24) | (-1.10) | |
| Oakland | 135 | -0.29 | 0.67 | 0.14 | 0.12 | 0.11 | -0.40 | 0.53 | 0.07 | -0.21 | 0.821 |
| | | (-2.23) | (10.98) | (2.04) | (1.80) | (1.84) | (-1.23) | (1.53) | (0.17) | (-0.56) | |
| | | | | | | | | | | | |
| | | | | | | | | | | | |
| Sacramento | 129 | -0.31 | 0.56 | 0.53 | -0.36 | 0.30 | 1.04 | 0.24 | 0.96 | -2.06 | 0.432 |
| | | (-0.97) | (3.11) | (2.81) | (-1.85) | (1.88) | (0.86) | (0.20) | (0.72) | (-2.02) | |
| Salinas | 135 | -0.22 | 0.71 | 0.30 | 0.04 | 0.04 | -0.34 | -0.34 | 0.060 | 0.42 | 0.831 |
| | | (-1.66) | (11.36) | (4.34) | (0.52) | (0.60) | (-1.02) | -(0.96) | (0.15) | (1.12) | |
| San Jose | 127 | -0.15 | 0.54 | 0.19 | 0.29 | -0.03 | -2.51 | 1.74 | 0.41 | -0.21 | 0.483 |
| | | (-0.56) | (3.53) | (0.99) | (1.52) | (-0.19) | (-2.43) | (1.36) | (0.31) | (-0.20) | |
| Santa Cruz | 133 | -0.26 | 0.42 | 0.39 | 0.07 | 0.10 | -1.20 | 0.77 | 0.69 | -0.85 | 0.646 |
| | | (-1.31) | (4.48) | (3.74) | (0.64) | (1.05) | (-2.27) | (1.28) | (1.12) | (-1.47) | |
| Santa Rosa | 131 | -0.67 | 0.46 | 0.43 | -0.04 | 0.25 | 3.29 | -1.25 | 0.74 | -1.56 | 0.457 |
| | | (-2.13) | (2.88) | (2.43) | (-0.21) | (1.58) | (3.27) | (-1.21) | (0.64) | (-1.58) | |
| Stockton | 127 | -0.59 | 0.54 | 0.12 | 0.37 | 0.02 | 0.30 | -0.33 | 1.58 | -1.07 | 0.456 |
| | | (-1.97) | (3.14) | (0.54) | (1.70) | (0.13) | (0.26) | (-0.23) | (1.06) | (-0.90) | |
| Vallejo | 117 | -0.92 | 0.36 | 0.72 | -0.04 | 0.10 | 0.88 | 0.81 | 2.46 | -2.35 | 0.292 |
| | | (-1.88) | (1.18) | (1.90) | (-0.10) | (0.34) | (0.41) | (0.31) | (0.96) | (-1.19) | |

## Panel C: Explanatory MSA - Santa Barbara

| | N | Constant | Lag0 | Lag1 | Lag2 | Lag3 | Lag0 | Lag1 | Lag2 | Lag3 | |
|---|---|---|---|---|---|---|---|---|---|---|---|
| Oxnard | 126 | 0.16 | 0.55 | 0.22 | 0.11 | 0.01 | -0.07 | 0.80 | 0.31 | -0.59 | 0.671 |
| | | (0.79) | (5.71) | (2.70) | (1.27) | (0.11) | (-0.09) | (1.29) | (0.53) | (-0.93) | |
| San Luis Obispo | 126 | -0.01 | 0.30 | 0.16 | 0.15 | 0.25 | -1.19 | -0.45 | 0.86 | 0.18 | 0.371 |
| | | (-0.04) | (2.03) | (1.29) | (1.17) | (1.68) | (-1.06) | (-0.47) | (0.95) | (0.19) | |

Notes: Regression results for a selection of California MSAs on contemporaneous and lagged returns (3 lags) of large coastal California MSAs. In addition the regressions contain four more variables, each one being an interaction between the explanatory city's return (including 3 lags) and a contemporaneous residual from a time trend fit of the log of the large coastal city’s house price index. The three large coastal leading cities used in the regressions are Los Angeles (Panel A), San Francesco (Panel B) and Santa Barbara (Panel C). N is the number of quarters in each regression. Regression coefficients and t-statistics in parentheses are given. R-squares of each regression are also reported. Results for some MSAs required a Cochrane-Orcutt adjustment for error term serial correlation. Durbin-Watson statistics for all presented MSA regressions allow us to reject the null hypothesis of first order serial correlation.

## Appendix Table 1
## Factor Model Data and Specification

| Data | Data Defined |
|---|---|
| MSA HP | log percent change in MSA house price index |
| CNP16OV | log percent change civilian non-institutional population |
| CPILFESL | log percent change in CPI |
| FEDFUNDS | log Fed Funds Rate |
| GS10 | log 10-year constant maturity Treasury |
| INDPRO | log percent change in Industrial Production Index |
| PAYEMS | log percent change in US payroll employment |
| PERMIT1 | log single-family building permits |
| PPIITM | log percent change PPI materials prices |
| UMCSENT | log University of Michigan Consumer Sentiment Index |
| UNRATE | log unemployment rate |
| SP500 | log percent change in S&P 500 |
| INCOME | log personal income |

Notes: MSA level data are quarterly and the start of the database is 1975 quarter 1 and the end is 2010 quarter 1. The number of MSAs in the database increases over time beginning with 2 in 1975 and reaches 380 by 1993. At the end of the sample there are 384 MSAs. All factor data are quarterly from 1975:Q1 – 2010:Q1 with the exception of UMCSENT which is available since 1977 quarter 4. The MSA house price data is provided by the Federal Housing Finance Agency (FHFA). MSA house price returns are computed as the log quarterly difference in the MSA repeat home sales price index. Data for the factors are obtained from the Federal Reserve Bank of St. Louis FRED (Federal Reserve Economic Data) except the SP500 (Datastream) and INCOME (US Dept of Commerce National Income and Product Accounts).

# Appendix Table 2
# Integration Details for All MSAs

| MSA | State | Mean | Rank Mean | Quintile Mean | Sigma | Rank Sigma | Quintile Sigma | Final R-Square | Change in R-Square | Trend t-stat | Rank Trend t-stat | Quintile Trend t-stat |
|---|---|---|---|---|---|---|---|---|---|---|---|---|
| Abilene | TX | 0.615 | 16 | 1 | 3.115 | 325 | 5 | 0.962 | 0.274 | -0.379 | 111 | 2 |
| Akron | OH | 0.957 | 200 | 3 | 1.688 | 68 | 1 | 0.897 | 0.049 | -1.964 | 52 | 1 |
| Albany | GA | 0.699 | 36 | 1 | 2.040 | 137 | 2 | 0.846 | 0.364 | -1.417 | 32 | 1 |
| Albany | NY | 1.337 | 341 | 5 | 2.590 | 241 | 4 | 0.871 | 0.134 | -2.847 | 67 | 1 |
| Albuquerque | NM | 1.152 | 298 | 4 | 1.961 | 116 | 2 | 0.938 | 0.062 | 0.783 | 171 | 3 |
| Alexandria | LA | 0.790 | 83 | 2 | 2.186 | 160 | 3 | 0.954 | 0.102 | 7.057 | 373 | 5 |
| Allentown | PA | 1.050 | 264 | 4 | 3.357 | 340 | 5 | 0.952 | 0.304 | -1.944 | 53 | 1 |
| Altoona | PA | 1.027 | 252 | 4 | 2.526 | 225 | 3 | 0.956 | 0.058 | -2.983 | 31 | 1 |
| Amarillo | TX | 0.779 | 73 | 1 | 2.894 | 298 | 4 | 0.734 | -0.012 | 0.174 | 136 | 2 |
| Ames | IA | 0.952 | 195 | 3 | 1.380 | 26 | 1 | 0.555 | 0.111 | -3.058 | 29 | 1 |
| Anchorage | AK | 0.728 | 51 | 1 | 3.811 | 360 | 5 | 0.777 | -0.047 | 2.316 | 249 | 4 |
| Anderson | SC | 0.829 | 110 | 2 | 2.131 | 41 | 1 | 0.851 | 0.197 | 1.453 | 203 | 3 |
| Anderson | IN | 0.872 | 140 | 2 | 1.513 | 150 | 2 | 0.944 | 0.576 | 6.308 | 366 | 5 |
| Ann Arbor | MI | 0.977 | 213 | 3 | 2.351 | 189 | 3 | 0.969 | 0.202 | 1.343 | 196 | 3 |
| Anniston | AL | 0.910 | 173 | 3 | 1.983 | 121 | 2 | 0.762 | 0.181 | 0.132 | 133 | 2 |
| Appleton | WI | 0.843 | 122 | 2 | 1.094 | 5 | 1 | 0.766 | 0.095 | 1.781 | 219 | 3 |
| Asheville | NC | 1.265 | 325 | 5 | 1.459 | 32 | 1 | 0.882 | 0.447 | 5.743 | 360 | 5 |
| Athens | GA | 0.904 | 165 | 3 | 1.255 | 12 | 1 | 0.814 | -0.104 | -2.158 | 46 | 1 |
| Atlanta | GA | 1.010 | 241 | 4 | 1.494 | 39 | 1 | 0.925 | 0.061 | -1.147 | 79 | 1 |
| Atlantic City | NJ | 1.176 | 304 | 4 | 2.407 | 203 | 3 | 0.954 | 0.356 | 1.468 | 204 | 3 |
| Auburn | AL | 0.858 | 131 | 2 | 2.223 | 167 | 3 | 0.965 | 0.574 | 6.963 | 372 | 5 |
| Augusta | GA | 0.876 | 143 | 2 | 3.216 | 332 | 5 | 0.967 | 0.383 | 0.653 | 159 | 3 |
| Austin | TX | 1.220 | 314 | 5 | 3.056 | 318 | 5 | 0.776 | 0.254 | 1.595 | 210 | 3 |
| Bakersfield | CA | 0.864 | 136 | 2 | 3.197 | 330 | 5 | 0.898 | 0.166 | 4.228 | 335 | 5 |
| Baltimore | MD | 1.364 | 349 | 5 | 1.846 | 103 | 2 | 0.750 | 0.035 | 2.692 | 272 | 4 |
| Bangor | ME | 0.721 | 46 | 1 | 2.700 | 261 | 4 | 0.964 | 0.440 | 3.558 | 316 | 5 |
| Barnstable Town | MA | 1.390 | 354 | 5 | 2.540 | 227 | 3 | 0.836 | 0.252 | 2.816 | 276 | 4 |
| Baton Rouge | LA | 0.915 | 177 | 3 | 1.704 | 70 | 1 | 0.786 | -0.095 | 2.360 | 257 | 4 |
| Battle Creek | MI | 0.897 | 159 | 3 | 2.136 | 151 | 2 | 0.855 | 0.342 | 2.565 | 269 | 4 |
| Bay City | MI | 0.899 | 161 | 3 | 2.513 | 221 | 3 | 0.860 | 0.243 | 1.847 | 221 | 3 |
| Beaumont | TX | 0.729 | 52 | 1 | 2.435 | 209 | 3 | 0.853 | 0.028 | -1.434 | 66 | 1 |
| Bellingham | WA | 1.337 | 342 | 5 | 2.565 | 235 | 4 | 0.743 | -0.080 | -3.500 | 20 | 1 |
| Bend | OR | 1.367 | 350 | 5 | 3.076 | 320 | 5 | 0.797 | 0.022 | 0.986 | 181 | 3 |
| Bethesda | MD | 1.438 | 361 | 5 | 2.232 | 168 | 3 | 0.760 | -0.038 | -0.973 | 83 | 2 |
| Billings | MT | 1.006 | 238 | 4 | 2.501 | 219 | 3 | 0.768 | 0.142 | -0.485 | 109 | 2 |
| Binghamton | NY | 0.809 | 101 | 2 | 2.516 | 223 | 3 | 0.739 | 0.102 | -2.385 | 41 | 1 |
| Birmingham | AL | 0.954 | 197 | 3 | 2.397 | 201 | 3 | 0.975 | 0.450 | 0.435 | 153 | 2 |
| Bismarck | ND | 0.967 | 205 | 3 | 1.349 | 21 | 1 | 0.666 | -0.177 | -6.109 | 2 | 1 |
| Blacksburg | VA | 0.996 | 227 | 3 | 1.519 | 42 | 1 | 0.716 | -0.081 | -1.746 | 57 | 1 |
| Bloomington | IN | 1.019 | 81 | 2 | 1.765 | 6 | 1 | 0.815 | 0.141 | 5.729 | 209 | 3 |
| Bloomington | IL | 0.788 | 247 | 4 | 1.108 | 82 | 2 | 0.787 | 0.043 | 1.593 | 359 | 5 |
| Boise City | ID | 0.849 | 126 | 2 | 3.337 | 338 | 5 | 0.896 | -0.039 | 3.439 | 308 | 5 |
| Boston | MA | 1.738 | 381 | 5 | 2.578 | 238 | 4 | 0.880 | 0.022 | 2.660 | 270 | 4 |
| Boulder | CO | 1.325 | 339 | 5 | 2.238 | 170 | 3 | 0.597 | -0.372 | -0.683 | 94 | 2 |
| Bowling Green | KY | 0.840 | 117 | 2 | 1.630 | 59 | 1 | 0.898 | 0.278 | -4.927 | 7 | 1 |
| Bremerton | WA | 1.187 | 308 | 5 | 2.817 | 280 | 4 | 0.914 | 0.039 | -2.616 | 37 | 1 |
| Bridgeport | CT | 1.428 | 360 | 5 | 2.703 | 262 | 4 | 0.857 | 0.098 | 1.992 | 227 | 3 |
| Brownsville | TX | 0.750 | 55 | 1 | 2.642 | 251 | 4 | 0.718 | -0.278 | -4.158 | 11 | 1 |
| Brunswick | GA | 1.162 | 302 | 4 | 1.966 | 118 | 2 | 0.817 | -0.096 | 0.653 | 160 | 3 |
| Buffalo | NY | 1.050 | 265 | 4 | 2.201 | 162 | 3 | 0.618 | -0.234 | 3.434 | 307 | 5 |
| Burlington | NC | 0.792 | 85 | 2 | 1.541 | 46 | 1 | 0.838 | 0.086 | 0.302 | 145 | 2 |
| Burlington | VT | 1.232 | 317 | 5 | 1.569 | 49 | 1 | 0.749 | 0.022 | 1.701 | 214 | 3 |
| Cambridge | MA | 1.712 | 379 | 5 | 2.386 | 196 | 3 | 0.671 | -0.261 | 0.720 | 165 | 3 |
| Camden | NJ | 1.358 | 348 | 5 | 2.466 | 214 | 3 | 0.800 | 0.204 | 3.116 | 293 | 4 |
| Canton | OH | 0.834 | 114 | 2 | 2.285 | 178 | 3 | 0.766 | -0.102 | -3.296 | 24 | 1 |
| Cape Coral | FL | 0.655 | 28 | 1 | 3.659 | 358 | 5 | 0.738 | 0.019 | 2.754 | 274 | 4 |
| Cape Girardeau | MO | 0.807 | 99 | 2 | 2.030 | 134 | 2 | 0.856 | 0.040 | -3.712 | 14 | 1 |
| Carson City | NV | 0.967 | 206 | 3 | 2.827 | 283 | 4 | 0.916 | 0.078 | 1.869 | 223 | 3 |
| Casper | WY | 0.727 | 49 | 1 | 4.553 | 375 | 5 | 0.835 | 0.198 | 0.360 | 149 | 2 |
| Cedar Rapids | IA | 0.842 | 119 | 2 | 2.035 | 136 | 2 | 0.757 | -0.119 | 3.680 | 320 | 5 |
| Champaign | IL | 0.833 | 112 | 2 | 1.279 | 14 | 1 | 0.897 | 0.397 | 5.474 | 354 | 5 |
| Charleston | WV | 0.755 | 60 | 1 | 1.800 | 90 | 2 | 0.865 | 0.378 | 0.954 | 178 | 3 |
| Charleston | SC | 1.254 | 322 | 5 | 5.619 | 381 | 5 | 0.957 | 0.189 | 7.573 | 377 | 5 |
| Charlotte | NC | 1.169 | 303 | 4 | 1.762 | 80 | 2 | 0.935 | 0.011 | 2.510 | 267 | 4 |
| Charlottesville | VA | 1.231 | 316 | 5 | 2.136 | 152 | 2 | 0.700 | -0.044 | -3.595 | 19 | 1 |
| Chattanooga | TN | 1.029 | 253 | 4 | 2.635 | 249 | 4 | 0.934 | 0.462 | -2.258 | 44 | 1 |
| Cheyenne | WY | 0.992 | 222 | 3 | 2.794 | 275 | 4 | 0.843 | 0.297 | 2.197 | 245 | 4 |
| Chicago | IL | 1.250 | 320 | 5 | 1.948 | 114 | 2 | 0.913 | 0.038 | 3.457 | 309 | 5 |
| Chico | CA | 1.066 | 273 | 4 | 3.077 | 321 | 5 | 0.832 | 0.169 | -0.844 | 87 | 2 |
| Cincinnati | OH | 0.997 | 228 | 3 | 1.206 | 10 | 1 | 0.953 | 0.695 | 3.113 | 292 | 4 |
| Clarksville | TN | 0.950 | 191 | 3 | 1.308 | 17 | 1 | 0.845 | 0.311 | 0.402 | 151 | 2 |
| Cleveland | TN | 0.983 | 203 | 3 | 1.900 | 109 | 2 | 0.898 | 0.084 | -1.074 | 80 | 2 |
| Cleveland | OH | 0.964 | 216 | 3 | 2.009 | 131 | 2 | 0.885 | 0.095 | 2.333 | 251 | 4 |
| Coeur d'Alene | ID | 1.298 | 333 | 5 | 2.756 | 271 | 4 | 0.813 | -0.110 | 3.304 | 301 | 4 |
| College Station | TX | 0.632 | 21 | 1 | 1.806 | 93 | 2 | 0.448 | -0.022 | 0.302 | 144 | 2 |
| Colorado Springs | CO | 1.046 | 261 | 4 | 2.590 | 242 | 4 | 0.946 | -0.004 | -0.257 | 115 | 2 |
| Columbia | SC | 0.765 | 68 | 1 | 1.413 | 30 | 1 | 0.764 | 0.121 | -1.292 | 72 | 1 |
| Columbia | MO | 0.994 | 223 | 3 | 1.669 | 65 | 1 | 0.957 | 0.320 | 1.381 | 199 | 3 |
| Columbus | OH | 0.817 | 103 | 2 | 1.258 | 11 | 1 | 0.845 | 0.072 | 1.199 | 113 | 2 |
| Columbus | GA | 0.943 | 186 | 3 | 1.254 | 13 | 1 | 0.915 | 0.283 | 2.103 | 188 | 3 |
| Columbus | IN | 0.995 | 225 | 3 | 1.491 | 38 | 1 | 0.960 | 0.090 | -0.331 | 236 | 4 |
| Corpus Christi | TX | 0.688 | 35 | 1 | 3.129 | 327 | 5 | 0.564 | -0.371 | -1.678 | 60 | 1 |
| Corvallis | OR | 1.305 | 335 | 5 | 2.222 | 166 | 3 | 0.913 | 0.190 | 2.860 | 278 | 4 |
| Crestview | FL | 0.940 | 184 | 3 | 3.041 | 317 | 5 | 0.690 | -0.238 | 4.394 | 337 | 5 |
| Cumberland | MD | 1.029 | 254 | 4 | 3.039 | 316 | 5 | 0.576 | 0.042 | 2.087 | 234 | 4 |
| Dallas | TX | 1.048 | 263 | 4 | 2.596 | 243 | 4 | 0.812 | 0.167 | -0.779 | 90 | 2 |
| Dalton | GA | 0.836 | 116 | 2 | 2.416 | 206 | 3 | 0.760 | 0.085 | -2.072 | 49 | 1 |
| Danville | IL | 0.730 | 53 | 1 | 2.446 | 200 | 3 | 0.931 | 0.304 | 4.740 | 27 | 1 |
| Danville | VA | 0.787 | 80 | 2 | 2.393 | 211 | 3 | 0.774 | -0.085 | -3.151 | 346 | 5 |
| Davenport | IA | 0.687 | 34 | 1 | 2.581 | 240 | 4 | 0.812 | -0.047 | 0.145 | 134 | 2 |
| Dayton | OH | 0.903 | 164 | 3 | 2.119 | 148 | 2 | 0.943 | 0.195 | 2.107 | 237 | 4 |
| Decatur | AL | 0.664 | 30 | 1 | 1.332 | 18 | 1 | 0.533 | -0.387 | -7.246 | 1 | 1 |
| Decatur | IL | 0.716 | 42 | 1 | 1.843 | 102 | 2 | 0.953 | 0.081 | -2.301 | 43 | 1 |
| Deltona | FL | 1.035 | 257 | 4 | 5.225 | 379 | 5 | 0.819 | 0.155 | 0.347 | 147 | 2 |
| Denver | CO | 1.345 | 344 | 5 | 1.824 | 98 | 2 | 0.948 | 0.389 | 6.511 | 369 | 5 |
| Des Moines | IA | 0.876 | 144 | 2 | 2.984 | 311 | 5 | 0.848 | 0.113 | 2.340 | 252 | 4 |
| Detroit | MI | 0.927 | 180 | 3 | 2.568 | 237 | 4 | 0.866 | 0.049 | 2.476 | 266 | 4 |
| Dothan | AL | 0.802 | 94 | 2 | 1.990 | 125 | 2 | 0.775 | 0.220 | 2.475 | 265 | 4 |
| Dover | DE | 0.940 | 185 | 3 | 2.274 | 176 | 3 | 0.712 | -0.004 | 4.168 | 332 | 5 |
| Dubuque | IA | 1.031 | 255 | 4 | 1.611 | 56 | 1 | 0.790 | -0.038 | -2.245 | 45 | 1 |
| Duluth | MN | 1.320 | 337 | 5 | 1.607 | 53 | 1 | 0.858 | 0.104 | 1.756 | 216 | 3 |
| Durham | NC | 0.985 | 218 | 3 | 2.376 | 194 | 3 | 0.976 | 0.111 | 0.551 | 158 | 3 |
| Eau Claire | WI | 1.061 | 269 | 4 | 1.820 | 97 | 2 | 0.945 | 0.163 | 3.184 | 297 | 4 |
| Edison | NJ | 1.489 | 368 | 5 | 2.335 | 187 | 3 | 0.753 | 0.013 | -0.223 | 117 | 2 |
| El Centro | CA | 0.607 | 11 | 1 | 4.240 | 370 | 5 | 0.912 | 0.114 | 2.365 | 258 | 4 |
| El Paso | TX | 0.706 | 39 | 1 | 2.167 | 158 | 3 | 0.937 | 0.300 | -0.949 | 85 | 2 |
| Elizabethtown | KY | 0.977 | 214 | 3 | 1.741 | 76 | 1 | 0.876 | 0.212 | 6.697 | 370 | 5 |
| Elkhart | IN | 0.760 | 64 | 1 | 1.600 | 52 | 1 | 0.637 | -0.096 | -5.684 | 3 | 1 |
| Elmira | NY | 0.681 | 33 | 1 | 2.876 | 294 | 4 | 0.417 | -0.348 | -3.711 | 15 | 1 |
| Erie | PA | 0.909 | 171 | 3 | 2.077 | 141 | 2 | 0.832 | 0.255 | 8.267 | 381 | 5 |
| Eugene | OR | 1.248 | 319 | 5 | 3.941 | 363 | 5 | 0.874 | 0.135 | 5.615 | 358 | 5 |
| Evansville | IN | 0.573 | 7 | 1 | 2.126 | 149 | 2 | 0.836 | 0.092 | 0.275 | 143 | 2 |
| Fairbanks | AK | 0.655 | 29 | 1 | 5.709 | 382 | 5 | 0.873 | 0.057 | -3.731 | 13 | 1 |
| Fargo | ND | 0.808 | 100 | 2 | 1.731 | 74 | 1 | 0.847 | 0.193 | 1.563 | 208 | 3 |
| Farmington | NM | 1.078 | 277 | 4 | 2.853 | 289 | 4 | 0.855 | 0.040 | 2.519 | 268 | 4 |
| Fayetteville | NC | 0.763 | 66 | 1 | 1.485 | 36 | 1 | 0.835 | -0.031 | 0.355 | 148 | 2 |
| Fayetteville | AR | 0.805 | 96 | 2 | 4.157 | 368 | 5 | 0.970 | 0.233 | 7.112 | 375 | 5 |
| Flagstaff | AZ | 1.287 | 329 | 5 | 2.754 | 270 | 4 | 0.794 | -0.075 | 2.174 | 242 | 4 |
| Flint | MI | 0.753 | 58 | 1 | 4.387 | 374 | 5 | 0.912 | 0.166 | 3.693 | 321 | 5 |
| Florence | SC | 0.851 | 61 | 1 | 1.402 | 29 | 1 | 0.730 | 0.338 | 0.746 | 168 | 3 |
| Florence | AL | 0.757 | 127 | 2 | 1.851 | 105 | 2 | 0.865 | 0.231 | 2.414 | 260 | 4 |
| Fond du Lac | WI | 1.002 | 234 | 4 | 2.027 | 133 | 2 | 0.738 | 0.011 | -1.549 | 63 | 1 |
| Fort Collins | CO | 1.200 | 309 | 5 | 3.222 | 333 | 5 | 0.975 | 0.526 | 6.259 | 365 | 5 |
| Fort Smith | AR | 0.797 | 89 | 2 | 2.326 | 183 | 3 | 0.847 | 0.115 | 3.089 | 290 | 4 |
| Fort Wayne | IN | 0.704 | 38 | 1 | 2.672 | 254 | 4 | 0.858 | 0.327 | 3.293 | 300 | 4 |
| Fort Worth | TX | 0.896 | 157 | 3 | 1.519 | 43 | 1 | 0.943 | -0.006 | -2.453 | 40 | 1 |
| Fresno | CA | 1.075 | 276 | 4 | 3.198 | 331 | 5 | 0.833 | -0.004 | 2.174 | 241 | 4 |
| Ft. Lauderdale | FL | 1.025 | 251 | 4 | 5.088 | 378 | 5 | 0.815 | 0.048 | 1.599 | 211 | 3 |
| Gadsden | AL | 0.978 | 215 | 3 | 1.842 | 101 | 2 | 0.813 | 0.020 | 0.708 | 164 | 3 |
| Gainesville | GA | 0.961 | 70 | 1 | 2.845 | 140 | 2 | 0.775 | 0.277 | 6.331 | 175 | 3 |
| Gainesville | FL | 0.772 | 202 | 3 | 2.072 | 288 | 4 | 0.779 | 0.042 | 0.879 | 367 | 5 |
| Gary | IN | 0.856 | 130 | 2 | 1.970 | 120 | 2 | 0.889 | 0.067 | 2.084 | 233 | 4 |
| Glens Falls | NY | 0.901 | 162 | 3 | 2.958 | 307 | 5 | 0.855 | 0.016 | -1.242 | 75 | 1 |
| Goldsboro | NC | 0.800 | 90 | 2 | 1.608 | 54 | 1 | 0.738 | 0.083 | 0.976 | 179 | 3 |
| Grand Forks | ND | 1.013 | 242 | 4 | 1.900 | 110 | 2 | 0.460 | -0.037 | -3.287 | 25 | 1 |
| Grand Junction | CO | 0.896 | 158 | 3 | 5.445 | 380 | 5 | 0.983 | 0.070 | 0.770 | 169 | 3 |
| Grand Rapids | MI | 0.881 | 150 | 2 | 2.374 | 193 | 3 | 0.929 | 0.109 | -0.681 | 96 | 2 |
| Great Falls | MT | 1.067 | 274 | 4 | 1.644 | 62 | 1 | 0.789 | -0.083 | 0.186 | 137 | 2 |
| Greeley | CO | 0.801 | 91 | 2 | 2.716 | 264 | 4 | 0.848 | 0.330 | 3.230 | 298 | 4 |
| Green Bay | WI | 0.866 | 137 | 2 | 1.081 | 2 | 1 | 0.857 | 0.039 | -0.187 | 119 | 2 |
| Greensboro | NC | 0.889 | 153 | 2 | 1.740 | 75 | 1 | 0.982 | 0.047 | -1.261 | 73 | 1 |
| Greenville | NC | 0.725 | 48 | 1 | 1.371 | 24 | 1 | 0.755 | -0.109 | 0.666 | 36 | 1 |
| Greenville | SC | 0.884 | 152 | 2 | 3.439 | 344 | 5 | 0.985 | 0.100 | -2.684 | 162 | 3 |
| Gulfport | MS | 1.022 | 248 | 4 | 2.533 | 226 | 3 | 0.658 | -0.166 | 1.406 | 201 | 3 |
| Hagerstown | MD | 0.952 | 196 | 3 | 2.379 | 195 | 3 | 0.850 | 0.285 | 0.660 | 161 | 3 |
| Hanford | CA | 0.909 | 172 | 3 | 3.098 | 324 | 5 | 0.619 | 0.226 | 4.120 | 331 | 5 |
| Harrisburg | PA | 1.006 | 239 | 4 | 3.261 | 335 | 5 | 0.934 | 0.588 | 3.974 | 329 | 5 |
| Harrisonburg | VA | 1.073 | 275 | 4 | 2.083 | 142 | 2 | 0.945 | 0.189 | 3.182 | 296 | 4 |
| Hartford | CT | 1.355 | 346 | 5 | 2.659 | 252 | 4 | 0.831 | 0.077 | 0.419 | 152 | 2 |
| Hattiesburg | MS | 0.792 | 86 | 2 | 2.262 | 175 | 3 | 0.884 | 0.102 | -0.090 | 125 | 2 |
| Hickory | NC | 0.933 | 181 | 3 | 1.435 | 31 | 1 | 0.935 | 0.393 | 2.348 | 254 | 4 |
| Hinesville | GA | 1.211 | 311 | 5 | 4.162 | 369 | 5 | 0.744 | 0.196 | -1.160 | 78 | 1 |
| Holland | MI | 0.955 | 198 | 3 | 2.232 | 169 | 3 | 0.942 | -0.012 | 3.628 | 317 | 5 |
| Honolulu | HI | 1.595 | 372 | 5 | 9.258 | 384 | 5 | 0.953 | 0.541 | 7.366 | 376 | 5 |
| Hot Springs | AR | 1.057 | 268 | 4 | 2.008 | 130 | 2 | 0.718 | -0.033 | 3.123 | 294 | 4 |
| Houma | LA | 0.919 | 178 | 3 | 2.608 | 247 | 4 | 0.715 | -0.100 | 5.211 | 351 | 5 |
| Houston | TX | 0.872 | 141 | 2 | 1.960 | 115 | 2 | 0.723 | -0.149 | 1.216 | 189 | 3 |
| Huntington | WV | 0.881 | 151 | 2 | 2.201 | 163 | 3 | 0.791 | 0.119 | -1.353 | 70 | 1 |
| Huntsville | AL | 0.776 | 71 | 1 | 0.980 | 1 | 1 | 0.968 | 0.200 | 3.545 | 314 | 5 |
| Idaho Falls | ID | 0.823 | 107 | 2 | 1.924 | 112 | 2 | 0.883 | 0.251 | -5.266 | 6 | 1 |
| Indianapolis | IN | 0.987 | 219 | 3 | 1.460 | 33 | 1 | 0.925 | 0.437 | 3.772 | 325 | 5 |
| Iowa City | IA | 0.946 | 188 | 3 | 1.636 | 61 | 1 | 0.779 | -0.005 | 2.072 | 232 | 4 |
| Ithaca | NY | 0.901 | 163 | 3 | 2.736 | 267 | 4 | 0.848 | 0.234 | -1.675 | 61 | 1 |
| Jackson | TN | 0.971 | 14 | 1 | 2.217 | 37 | 1 | 0.879 | 0.553 | 4.549 | 8 | 1 |
| Jackson | MI | 0.612 | 19 | 1 | 3.510 | 165 | 3 | 0.958 | 0.239 | 3.866 | 327 | 5 |
| Jackson | MS | 0.625 | 209 | 3 | 1.490 | 347 | 5 | 0.884 | 0.134 | -4.466 | 341 | 5 |
| Jacksonville | NC | 1.080 | 278 | 4 | 2.249 | 173 | 3 | 0.941 | 0.117 | 3.549 | 264 | 4 |
| Jacksonville | FL | 1.131 | 292 | 4 | 2.623 | 248 | 4 | 0.936 | 0.202 | 2.457 | 315 | 5 |
| Janesville | WI | 0.895 | 156 | 3 | 1.471 | 35 | 1 | 0.925 | 0.448 | 2.681 | 271 | 4 |
| Jefferson City | MO | 0.795 | 87 | 2 | 1.561 | 48 | 1 | 0.828 | 0.114 | 3.368 | 302 | 4 |
| Johnson City | TN | 1.062 | 270 | 4 | 1.670 | 67 | 1 | 0.972 | 0.523 | 5.340 | 352 | 5 |
| Johnstown | PA | 0.834 | 115 | 2 | 2.691 | 258 | 4 | 0.529 | -0.093 | -4.440 | 9 | 1 |
| Jonesboro | AR | 0.675 | 32 | 1 | 1.886 | 107 | 2 | 0.756 | 0.085 | 0.865 | 174 | 3 |
| Joplin | MO | 0.713 | 41 | 1 | 1.540 | 45 | 1 | 0.784 | 0.291 | -3.193 | 26 | 1 |
| Kalamazoo | MI | 0.951 | 193 | 3 | 2.023 | 132 | 2 | 0.940 | 0.047 | 5.508 | 356 | 5 |
| Kankakee | IL | 1.130 | 291 | 4 | 1.850 | 104 | 2 | 0.862 | -0.059 | 2.347 | 253 | 4 |
| Kansas City | MO | 0.970 | 208 | 3 | 1.653 | 64 | 1 | 0.914 | 0.350 | 1.539 | 207 | 3 |
| Kennewick | WA | 1.000 | 232 | 4 | 3.631 | 357 | 5 | 0.713 | -0.180 | -0.297 | 114 | 2 |
| Killeen | TX | 0.582 | 8 | 1 | 2.543 | 230 | 3 | 0.790 | -0.083 | -5.326 | 5 | 1 |
| Kingsport | TN | 0.951 | 194 | 3 | 1.841 | 100 | 2 | 0.791 | 0.133 | 0.525 | 156 | 3 |
| Kingston | NY | 1.018 | 246 | 4 | 2.928 | 303 | 4 | 0.753 | 0.010 | 2.195 | 244 | 4 |
| Knoxville | TN | 0.906 | 167 | 3 | 1.133 | 7 | 1 | 0.840 | 0.083 | -3.359 | 22 | 1 |
| Kokomo | IN | 0.585 | 10 | 1 | 2.147 | 155 | 2 | 0.484 | -0.139 | -2.471 | 39 | 1 |
| La Crosse | WI | 0.987 | 220 | 3 | 1.177 | 8 | 1 | 0.817 | 0.006 | 3.754 | 324 | 5 |
| Lafayette | LA | 0.817 | 15 | 1 | 1.190 | 9 | 1 | 0.644 | -0.182 | 1.264 | 190 | 3 |
| Lafayette | IN | 0.612 | 104 | 2 | 2.559 | 233 | 4 | 0.629 | -0.119 | 2.022 | 230 | 3 |
| Lake Charles | LA | 0.938 | 183 | 3 | 2.412 | 204 | 3 | 0.895 | 0.013 | -0.607 | 103 | 2 |
| Lake County | IL | 1.005 | 235 | 4 | 2.143 | 154 | 2 | 0.941 | 0.095 | 3.017 | 287 | 4 |
| Lake Havasu City | AZ | 0.781 | 76 | 1 | 3.162 | 328 | 5 | 0.862 | 0.130 | 5.943 | 363 | 5 |
| Lakeland | FL | 0.716 | 43 | 1 | 2.759 | 272 | 4 | 0.847 | 0.100 | 6.884 | 371 | 5 |
| Lancaster | PA | 0.995 | 226 | 3 | 2.459 | 212 | 3 | 0.947 | 0.132 | 2.834 | 277 | 4 |
| Lansing | MI | 0.851 | 128 | 2 | 2.968 | 309 | 5 | 0.864 | -0.078 | -1.301 | 71 | 1 |
| Laredo | TX | 0.666 | 31 | 1 | 2.826 | 282 | 4 | 0.847 | -0.148 | 0.216 | 139 | 2 |
| Las Cruces | NM | 0.823 | 108 | 2 | 1.839 | 99 | 2 | 0.824 | 0.101 | 3.516 | 313 | 5 |
| Las Vegas | NV | 0.750 | 56 | 1 | 4.303 | 372 | 5 | 0.735 | 0.240 | 4.830 | 348 | 5 |
| Lawrence | KS | 1.054 | 267 | 4 | 1.784 | 85 | 2 | 0.876 | 0.147 | 2.928 | 280 | 4 |
| Lawton | OK | 0.806 | 97 | 2 | 2.549 | 232 | 4 | 0.849 | 0.054 | 2.412 | 259 | 4 |
| Lebanon | PA | 0.997 | 229 | 3 | 1.999 | 128 | 2 | 0.676 | 0.343 | -2.508 | 38 | 1 |
| Lewiston | ID | 1.286 | 175 | 3 | 2.094 | 145 | 2 | 0.632 | 0.166 | 0.127 | 131 | 2 |
| Lewiston | ME | 0.914 | 328 | 5 | 2.696 | 259 | 4 | 0.876 | 0.108 | 0.775 | 170 | 3 |
| Lexington | KY | 0.795 | 88 | 2 | 2.690 | 257 | 4 | 0.855 | 0.410 | -0.675 | 99 | 2 |
| Lima | OH | 0.716 | 44 | 1 | 1.962 | 117 | 2 | 0.813 | 0.097 | 4.578 | 342 | 5 |
| Lincoln | NE | 0.801 | 92 | 2 | 1.669 | 66 | 1 | 0.918 | 0.091 | 0.500 | 154 | 2 |
| Little Rock | AR | 0.949 | 189 | 3 | 2.469 | 215 | 3 | 0.826 | 0.376 | 1.046 | 184 | 3 |
| Logan | UT | 1.063 | 271 | 4 | 1.988 | 124 | 2 | 0.724 | 0.099 | -0.679 | 97 | 2 |
| Longview | WA | 0.644 | 25 | 1 | 2.661 | 253 | 4 | 0.727 | -0.231 | 4.030 | 330 | 5 |
| Longview | TX | 1.023 | 249 | 4 | 5.746 | 383 | 5 | 0.966 | 0.100 | 6.257 | 364 | 5 |
| Los Angeles | CA | 1.736 | 380 | 5 | 2.839 | 286 | 4 | 0.558 | -0.339 | 2.172 | 239 | 4 |
| Louisville | KY | 1.122 | 289 | 4 | 1.350 | 22 | 1 | 0.955 | 0.190 | 1.296 | 193 | 3 |
| Lubbock | TX | 0.607 | 12 | 1 | 2.579 | 239 | 4 | 0.955 | 0.124 | -3.671 | 17 | 1 |
| Lynchburg | VA | 1.024 | 250 | 4 | 1.578 | 51 | 1 | 0.958 | 0.215 | 5.011 | 349 | 5 |
| Macon | GA | 0.908 | 169 | 3 | 2.705 | 263 | 4 | 0.922 | -0.034 | -1.708 | 59 | 1 |
| Madera | CA | 0.879 | 146 | 2 | 3.548 | 351 | 5 | 0.826 | -0.121 | 8.208 | 380 | 5 |
| Madison | WI | 1.115 | 286 | 4 | 2.327 | 184 | 3 | 0.867 | 0.074 | -0.056 | 127 | 2 |
| Manchester | NH | 1.214 | 313 | 5 | 2.420 | 207 | 3 | 0.918 | 0.051 | -1.910 | 54 | 1 |
| Manhattan | KS | 1.032 | 256 | 4 | 2.833 | 285 | 4 | 0.755 | -0.013 | -3.702 | 16 | 1 |
| Mankato | MN | 1.007 | 240 | 4 | 1.787 | 87 | 2 | 0.937 | 0.053 | 0.982 | 180 | 3 |
| Mansfield | OH | 0.770 | 69 | 1 | 2.248 | 172 | 3 | 0.636 | 0.025 | -0.337 | 112 | 2 |
| McAllen | TX | 0.541 | 4 | 1 | 3.393 | 341 | 5 | 0.921 | 0.202 | 1.472 | 205 | 3 |
| Medford | OR | 1.177 | 305 | 4 | 3.127 | 326 | 5 | 0.933 | 0.276 | 3.274 | 299 | 4 |
| Memphis | TN | 0.889 | 154 | 2 | 3.081 | 322 | 5 | 0.943 | 0.058 | 1.716 | 215 | 3 |
| Merced | CA | 0.790 | 84 | 2 | 4.674 | 376 | 5 | 0.889 | 0.111 | 2.937 | 282 | 4 |
| Miami | FL | 1.308 | 336 | 5 | 3.823 | 361 | 5 | 0.935 | 0.106 | -0.133 | 124 | 2 |
| Michigan City | IN | 1.064 | 272 | 4 | 2.373 | 192 | 3 | 0.898 | 0.064 | 7.098 | 374 | 5 |
| Midland | TX | 0.493 | 2 | 1 | 2.869 | 293 | 4 | 0.742 | -0.083 | 4.516 | 340 | 5 |
| Milwaukee | WI | 1.047 | 262 | 4 | 1.755 | 79 | 1 | 0.945 | 0.508 | 5.614 | 357 | 5 |
| Minneapolis | MN | 1.213 | 312 | 5 | 1.814 | 95 | 2 | 0.900 | -0.025 | -0.616 | 102 | 2 |
| Missoula | MT | 1.439 | 362 | 5 | 2.860 | 292 | 4 | 0.993 | 0.316 | -0.682 | 95 | 2 |
| Mobile | AL | 0.855 | 129 | 2 | 2.844 | 287 | 4 | 0.956 | 0.217 | 1.288 | 192 | 3 |
| Modesto | CA | 1.005 | 236 | 4 | 4.006 | 364 | 5 | 0.820 | 0.168 | 2.994 | 286 | 4 |
| Monroe | MI | 0.780 | 75 | 1 | 1.852 | 106 | 2 | 0.797 | 0.227 | 0.240 | 140 | 2 |
| Monroe | LA | 0.959 | 201 | 3 | 2.726 | 266 | 4 | 0.930 | 0.136 | 3.468 | 312 | 5 |
| Montgomery | AL | 0.617 | 17 | 1 | 1.283 | 15 | 1 | 0.853 | 0.159 | -0.894 | 86 | 2 |
| Morgantown | WV | 0.998 | 230 | 3 | 2.333 | 186 | 3 | 0.645 | -0.182 | 2.098 | 235 | 4 |
| Morristown | TN | 0.937 | 182 | 3 | 1.609 | 55 | 1 | 0.775 | 0.032 | 0.502 | 155 | 2 |
| Mount Vernon | WA | 1.424 | 357 | 5 | 3.485 | 346 | 5 | 0.940 | 0.003 | 0.380 | 150 | 2 |
| Muncie | IN | 0.553 | 5 | 1 | 2.007 | 129 | 2 | 0.622 | -0.107 | -5.328 | 4 | 1 |
| Muskegon | MI | 0.817 | 105 | 2 | 1.697 | 69 | 1 | 0.867 | 0.142 | -1.178 | 77 | 1 |
| Myrtle Beach | SC | 0.879 | 147 | 2 | 2.174 | 159 | 3 | 0.878 | -0.045 | -1.896 | 56 | 1 |
| Napa | CA | 1.424 | 358 | 5 | 2.989 | 312 | 5 | 0.838 | 0.158 | 3.463 | 310 | 5 |
| Naples | FL | 0.956 | 199 | 3 | 3.544 | 350 | 5 | 0.649 | -0.318 | 1.046 | 183 | 3 |
| Nashville | TN | 1.042 | 259 | 4 | 1.553 | 47 | 1 | 0.963 | 0.234 | 0.072 | 130 | 2 |
| Nassau | NY | 1.680 | 377 | 5 | 2.389 | 198 | 3 | 0.798 | 0.144 | 0.532 | 157 | 3 |
| New Haven | CT | 1.399 | 355 | 5 | 2.674 | 255 | 4 | 0.832 | 0.211 | -0.584 | 104 | 2 |
| New Orleans | LA | 1.087 | 279 | 4 | 2.153 | 156 | 3 | 0.855 | 0.292 | 2.428 | 263 | 4 |
| New York | NY | 1.673 | 375 | 5 | 2.428 | 208 | 3 | 0.638 | -0.085 | 1.319 | 195 | 3 |
| Newark | NJ | 1.574 | 370 | 5 | 2.321 | 181 | 3 | 0.677 | -0.166 | 1.994 | 228 | 3 |
| Niles | MI | 1.184 | 307 | 5 | 1.762 | 81 | 2 | 0.757 | -0.168 | -0.571 | 106 | 2 |
| North Port | FL | 1.014 | 243 | 4 | 4.257 | 371 | 5 | 0.699 | -0.084 | 3.639 | 318 | 5 |
| Norwich | CT | 1.001 | 233 | 4 | 2.279 | 177 | 3 | 0.887 | 0.342 | 0.130 | 132 | 2 |
| Oakland | CA | 1.699 | 378 | 5 | 2.638 | 250 | 4 | 0.577 | -0.115 | 0.744 | 167 | 3 |
| Ocala | FL | 0.722 | 47 | 1 | 2.808 | 278 | 4 | 0.776 | 0.103 | -0.071 | 126 | 2 |
| Ocean City | NJ | 1.477 | 366 | 5 | 2.829 | 284 | 4 | 0.831 | -0.028 | 3.466 | 311 | 5 |
| Odessa | TX | 0.430 | 1 | 1 | 3.843 | 362 | 5 | 0.726 | -0.099 | 5.413 | 353 | 5 |
| Ogden | UT | 0.974 | 212 | 3 | 2.931 | 304 | 4 | 0.952 | 0.167 | 3.948 | 328 | 5 |
| Oklahoma City | OK | 0.944 | 187 | 3 | 2.203 | 164 | 3 | 0.851 | 0.194 | 1.668 | 213 | 3 |
| Olympia | WA | 1.290 | 330 | 5 | 3.530 | 349 | 5 | 0.979 | 0.112 | 1.613 | 212 | 3 |
| Omaha | NE | 0.904 | 166 | 3 | 2.597 | 244 | 4 | 0.873 | 0.154 | 1.298 | 194 | 3 |
| Orlando | FL | 1.106 | 284 | 4 | 2.682 | 256 | 4 | 0.798 | 0.245 | -0.751 | 92 | 2 |
| Oshkosh | WI | 0.846 | 123 | 2 | 1.332 | 19 | 1 | 0.912 | 0.313 | 2.979 | 283 | 4 |
| Owensboro | KY | 0.727 | 50 | 1 | 2.289 | 179 | 3 | 0.697 | -0.101 | 2.932 | 281 | 4 |
| Oxnard | CA | 1.635 | 374 | 5 | 2.991 | 313 | 5 | 0.768 | 0.099 | 2.353 | 255 | 4 |
| Palm Bay | FL | 0.830 | 111 | 2 | 4.069 | 365 | 5 | 0.821 | -0.015 | 4.499 | 339 | 5 |
| Palm Coast | FL | 0.619 | 18 | 1 | 3.604 | 355 | 5 | 0.816 | -0.163 | -0.580 | 105 | 2 |
| Panama City | FL | 0.919 | 179 | 3 | 2.895 | 300 | 4 | 0.856 | 0.179 | 8.038 | 379 | 5 |
| Parkersburg | WV | 0.847 | 124 | 2 | 1.901 | 111 | 2 | 0.872 | 0.147 | -0.618 | 101 | 2 |
| Pascagoula | MS | 1.087 | 280 | 4 | 2.196 | 161 | 3 | 0.747 | 0.115 | 3.391 | 304 | 4 |
| Peabody | MA | 1.480 | 367 | 5 | 2.485 | 218 | 3 | 0.856 | 0.141 | -1.394 | 69 | 1 |
| Pensacola | FL | 0.860 | 132 | 2 | 2.247 | 171 | 3 | 0.798 | -0.036 | 2.014 | 229 | 3 |
| Peoria | IL | 0.641 | 23 | 1 | 4.077 | 366 | 5 | 0.900 | 0.175 | 9.310 | 383 | 5 |
| Philadelphia | PA | 1.343 | 343 | 5 | 1.742 | 77 | 1 | 0.749 | -0.113 | 2.172 | 240 | 4 |
| Phoenix | AZ | 1.104 | 283 | 4 | 2.979 | 310 | 5 | 0.726 | -0.026 | 4.721 | 345 | 5 |
| Pine Bluff | AR | 0.785 | 79 | 1 | 2.346 | 188 | 3 | 0.774 | 0.207 | 2.224 | 246 | 4 |
| Pittsburgh | PA | 1.005 | 237 | 4 | 1.998 | 127 | 2 | 0.957 | 0.409 | 1.779 | 218 | 3 |
| Pittsfield | MA | 0.898 | 160 | 3 | 2.937 | 306 | 4 | 0.970 | 0.051 | 2.360 | 256 | 4 |
| Pocatello | ID | 1.116 | 287 | 4 | 1.745 | 78 | 1 | 0.968 | 0.170 | 0.732 | 166 | 3 |
| Port St. Lucie | FL | 0.783 | 77 | 1 | 4.322 | 373 | 5 | 0.924 | -0.028 | 3.061 | 288 | 4 |
| Portland | ME | 1.356 | 347 | 5 | 2.083 | 143 | 2 | 0.853 | 0.075 | 1.972 | 226 | 3 |
| Portland | OR | 1.457 | 363 | 5 | 2.386 | 197 | 3 | 0.862 | 0.067 | 4.181 | 334 | 5 |
| Poughkeepsie | NY | 1.252 | 321 | 5 | 3.027 | 315 | 5 | 0.947 | 0.220 | -0.733 | 93 | 2 |
| Prescott | AZ | 1.036 | 258 | 4 | 2.566 | 236 | 4 | 0.916 | 0.331 | 1.124 | 186 | 3 |
| Providence | RI | 1.476 | 365 | 5 | 2.753 | 269 | 4 | 0.915 | 0.142 | -0.181 | 120 | 2 |
| Provo | UT | 0.990 | 221 | 3 | 2.402 | 202 | 3 | 0.960 | 0.308 | 10.469 | 384 | 5 |
| Pueblo | CO | 0.803 | 95 | 2 | 4.139 | 367 | 5 | 0.925 | 0.377 | 3.162 | 295 | 4 |
| Punta Gorda | FL | 0.759 | 62 | 1 | 3.562 | 353 | 5 | 0.817 | 0.072 | 3.738 | 323 | 5 |
| Racine | WI | 0.965 | 204 | 3 | 1.374 | 25 | 1 | 0.865 | 0.083 | 2.986 | 285 | 4 |
| Raleigh | NC | 1.098 | 281 | 4 | 1.628 | 58 | 1 | 0.949 | -0.008 | 3.406 | 305 | 4 |
| Rapid City | SD | 1.245 | 318 | 5 | 1.775 | 83 | 2 | 0.864 | 0.117 | -2.758 | 34 | 1 |
| Reading | PA | 0.983 | 217 | 3 | 2.523 | 224 | 3 | 0.964 | 0.340 | -1.570 | 62 | 1 |
| Redding | CA | 0.879 | 148 | 2 | 3.063 | 319 | 5 | 0.947 | 0.388 | -1.395 | 68 | 1 |
| Reno | NV | 0.789 | 82 | 2 | 2.907 | 302 | 4 | 0.834 | 0.296 | 0.336 | 146 | 2 |
| Richmond | VA | 1.149 | 297 | 4 | 1.783 | 84 | 2 | 0.889 | 0.226 | 0.861 | 173 | 3 |
| Riverside | CA | 1.296 | 332 | 5 | 3.438 | 343 | 5 | 0.713 | 0.177 | -1.260 | 74 | 1 |
| Roanoke | VA | 1.384 | 353 | 5 | 3.614 | 356 | 5 | 0.927 | 0.215 | 6.352 | 368 | 5 |
| Rochester | NY | 0.825 | 109 | 2 | 1.344 | 20 | 1 | 0.848 | 0.104 | 2.182 | 35 | 1 |
| Rochester | MN | 0.863 | 134 | 2 | 1.784 | 86 | 2 | 0.671 | -0.001 | -2.716 | 243 | 4 |
| Rockford | IL | 0.763 | 67 | 1 | 2.540 | 228 | 3 | 0.875 | 0.220 | 2.712 | 273 | 4 |
| Rockingham County | NH | 1.183 | 306 | 4 | 2.320 | 180 | 3 | 0.943 | 0.066 | -0.379 | 110 | 2 |
| Rocky Mount | NC | 0.630 | 20 | 1 | 1.790 | 88 | 2 | 0.776 | -0.054 | 1.393 | 200 | 3 |
| Rome | GA | 0.848 | 125 | 2 | 2.439 | 210 | 3 | 0.696 | 0.112 | 1.807 | 220 | 3 |
| Sacramento | CA | 1.354 | 345 | 5 | 2.894 | 299 | 4 | 0.649 | -0.176 | 2.980 | 284 | 4 |
| Saginaw | MI | 0.637 | 22 | 1 | 1.986 | 123 | 2 | 0.749 | 0.102 | 4.324 | 336 | 5 |
| Salem | OR | 1.138 | 294 | 4 | 2.771 | 273 | 4 | 0.934 | 0.582 | 1.920 | 224 | 3 |
| Salinas | CA | 1.336 | 340 | 5 | 3.570 | 354 | 5 | 0.613 | 0.044 | 1.866 | 222 | 3 |
| Salisbury | MD | 0.998 | 231 | 4 | 2.372 | 191 | 3 | 0.349 | -0.136 | 1.415 | 202 | 3 |
| Salt Lake City | UT | 1.274 | 327 | 5 | 2.413 | 205 | 3 | 0.968 | 0.121 | 3.851 | 326 | 5 |
| San Angelo | TX | 0.740 | 54 | 1 | 2.804 | 277 | 4 | 0.783 | -0.017 | -1.504 | 65 | 1 |
| San Antonio | TX | 0.784 | 78 | 1 | 3.354 | 339 | 5 | 0.866 | 0.266 | 5.040 | 350 | 5 |
| San Diego | CA | 1.541 | 369 | 5 | 3.013 | 314 | 5 | 0.868 | -0.114 | 0.253 | 141 | 2 |
| San Francisco | CA | 1.892 | 384 | 5 | 2.540 | 229 | 3 | 0.638 | -0.143 | -2.069 | 50 | 1 |
| San Jose | CA | 1.877 | 383 | 5 | 2.789 | 274 | 4 | 0.759 | 0.123 | -2.379 | 42 | 1 |
| San Luis Obispo | CA | 1.303 | 334 | 5 | 3.326 | 337 | 5 | 0.637 | -0.134 | -0.826 | 88 | 2 |
| Sandusky | OH | 0.863 | 135 | 2 | 2.855 | 290 | 4 | 0.853 | 0.023 | 3.731 | 322 | 5 |
| Santa Ana | CA | 1.674 | 376 | 5 | 2.718 | 265 | 4 | 0.626 | 0.051 | -1.713 | 58 | 1 |
| Santa Barbara | CA | 1.470 | 364 | 5 | 2.879 | 295 | 4 | 0.779 | 0.090 | 0.256 | 142 | 2 |
| Santa Cruz | CA | 1.599 | 373 | 5 | 3.093 | 323 | 5 | 0.657 | 0.065 | 2.422 | 262 | 4 |
| Santa Fe | NM | 1.101 | 282 | 4 | 2.032 | 135 | 2 | 0.819 | 0.195 | 3.650 | 319 | 5 |
| Santa Rosa | CA | 1.590 | 371 | 5 | 2.855 | 291 | 4 | 0.678 | 0.251 | 0.990 | 182 | 3 |
| Savannah | GA | 1.223 | 315 | 5 | 2.513 | 222 | 3 | 0.932 | 0.295 | 2.239 | 247 | 4 |
| Scranton | PA | 1.264 | 324 | 5 | 2.351 | 190 | 3 | 0.885 | 0.252 | 1.373 | 198 | 3 |
| Seattle | WA | 1.744 | 382 | 5 | 2.508 | 220 | 3 | 0.925 | 0.271 | 2.038 | 231 | 4 |
| Sebastian | FL | 0.711 | 40 | 1 | 3.554 | 352 | 5 | 0.770 | 0.108 | 4.677 | 343 | 5 |
| Sheboygan | WI | 0.949 | 190 | 3 | 1.390 | 27 | 1 | 0.845 | 0.019 | 2.152 | 238 | 4 |
| Sherman | TX | 0.533 | 3 | 1 | 2.887 | 297 | 4 | 0.843 | 0.031 | -2.842 | 33 | 1 |
| Shreveport | LA | 0.582 | 9 | 1 | 1.711 | 71 | 1 | 0.865 | 0.006 | 0.159 | 135 | 2 |
| Sioux City | IA | 0.910 | 174 | 3 | 1.721 | 73 | 1 | 0.618 | -0.170 | -0.643 | 100 | 2 |
| Sioux Falls | SD | 0.894 | 155 | 2 | 1.810 | 94 | 2 | 0.834 | 0.152 | -0.559 | 107 | 2 |
| South Bend | IN | 0.876 | 145 | 2 | 1.353 | 23 | 1 | 0.866 | 0.194 | 1.774 | 217 | 3 |
| Spartanburg | SC | 0.801 | 93 | 2 | 1.391 | 28 | 1 | 0.832 | 0.105 | 3.112 | 291 | 4 |
| Spokane | WA | 1.159 | 300 | 4 | 2.696 | 260 | 4 | 0.877 | -0.095 | 1.088 | 185 | 3 |
| Springfield | IL | 0.702 | 37 | 1 | 1.089 | 3 | 1 | 0.897 | 0.057 | 8.489 | 12 | 1 |
| Springfield | MA | 1.293 | 63 | 1 | 2.605 | 4 | 1 | 0.852 | 0.088 | 4.770 | 129 | 2 |
| Springfield | OH | 0.759 | 138 | 2 | 1.082 | 217 | 3 | 0.888 | 0.032 | 0.067 | 347 | 5 |
| Springfield | MO | 0.870 | 331 | 5 | 2.475 | 246 | 4 | 0.717 | -0.069 | -3.873 | 382 | 5 |
| St. Cloud | MN | 0.971 | 210 | 3 | 1.575 | 50 | 1 | 0.748 | -0.091 | -0.676 | 98 | 2 |
| St. George | UT | 0.842 | 120 | 2 | 3.517 | 348 | 5 | 0.835 | 0.005 | 1.517 | 206 | 3 |
| St. Joseph | MO | 1.017 | 245 | 4 | 1.994 | 126 | 2 | 0.545 | -0.404 | 1.933 | 225 | 3 |
| St. Louis | MO | 1.139 | 295 | 4 | 3.304 | 336 | 5 | 0.927 | 0.138 | 7.950 | 378 | 5 |
| State College | PA | 0.950 | 192 | 3 | 1.650 | 63 | 1 | 0.635 | -0.113 | -1.534 | 64 | 1 |
| Steubenville | WV | 0.777 | 72 | 1 | 2.881 | 296 | 4 | 0.638 | -0.038 | 4.716 | 344 | 5 |
| Stockton | CA | 1.050 | 266 | 4 | 3.696 | 359 | 5 | 0.669 | 0.238 | -0.537 | 108 | 2 |
| Sumter | SC | 0.842 | 121 | 2 | 1.805 | 92 | 2 | 0.816 | -0.046 | -1.999 | 51 | 1 |
| Syracuse | NY | 1.014 | 244 | 4 | 1.793 | 89 | 2 | 0.786 | 0.214 | -3.324 | 23 | 1 |
| Tacoma | WA | 1.425 | 359 | 5 | 2.460 | 213 | 3 | 0.929 | 0.035 | -0.210 | 118 | 2 |
| Tallahassee | FL | 0.908 | 170 | 3 | 2.328 | 185 | 3 | 0.914 | 0.180 | 2.861 | 279 | 4 |
| Tampa | FL | 1.117 | 288 | 4 | 2.597 | 245 | 4 | 0.744 | -0.190 | -2.146 | 48 | 1 |
| Terre Haute | IN | 0.643 | 24 | 1 | 2.139 | 153 | 2 | 0.764 | 0.238 | -0.785 | 89 | 2 |
| Texarkana | TX | 0.779 | 74 | 1 | 1.713 | 72 | 1 | 0.697 | -0.170 | -4.316 | 10 | 1 |
| Toledo | OH | 0.717 | 45 | 1 | 2.934 | 305 | 4 | 0.965 | 0.439 | 2.414 | 261 | 4 |
| Topeka | KS | 0.754 | 59 | 1 | 1.815 | 96 | 2 | 0.823 | 0.084 | 2.331 | 250 | 4 |
| Trenton | NJ | 1.324 | 338 | 5 | 2.470 | 216 | 3 | 0.834 | 0.242 | 0.910 | 176 | 3 |
| Tucson | AZ | 1.203 | 310 | 5 | 4.682 | 377 | 5 | 0.844 | -0.037 | 0.704 | 163 | 3 |
| Tulsa | OK | 0.907 | 168 | 3 | 1.969 | 119 | 2 | 0.799 | 0.243 | 0.804 | 172 | 3 |
| Tuscaloosa | AL | 0.994 | 224 | 3 | 1.293 | 16 | 1 | 0.954 | 0.140 | -0.768 | 91 | 2 |
| Tyler | TX | 0.568 | 6 | 1 | 2.087 | 144 | 2 | 0.592 | -0.290 | -1.031 | 81 | 2 |
| Utica | NY | 0.840 | 118 | 2 | 2.389 | 199 | 3 | 0.653 | -0.304 | -0.952 | 84 | 2 |
| Valdosta | GA | 0.914 | 176 | 3 | 2.071 | 139 | 2 | 0.855 | -0.010 | -0.143 | 123 | 2 |
| Vallejo | CA | 1.133 | 293 | 4 | 3.419 | 342 | 5 | 0.796 | 0.074 | -2.146 | 47 | 1 |
| Victoria | TX | 0.760 | 65 | 1 | 2.063 | 138 | 2 | 0.787 | -0.107 | -0.237 | 116 | 2 |
| Vineland | NJ | 1.146 | 296 | 4 | 2.821 | 281 | 4 | 0.886 | -0.052 | -3.110 | 28 | 1 |
| Virginia Beach | VA | 1.379 | 352 | 5 | 1.985 | 122 | 2 | 0.733 | -0.085 | 0.214 | 138 | 2 |
| Visalia | CA | 0.872 | 142 | 2 | 3.244 | 334 | 5 | 0.828 | -0.013 | 3.380 | 303 | 4 |
| Waco | TX | 0.610 | 13 | 1 | 2.109 | 147 | 2 | 0.357 | -0.616 | -1.185 | 76 | 1 |
| Warner Robins | GA | 0.644 | 26 | 1 | 1.465 | 34 | 1 | 0.682 | -0.018 | 2.302 | 248 | 4 |
| Warren | MI | 0.968 | 207 | 3 | 2.322 | 182 | 3 | 0.927 | 0.110 | 1.372 | 197 | 3 |
| Washington | DC | 1.411 | 356 | 5 | 2.101 | 146 | 2 | 0.569 | -0.025 | -3.617 | 18 | 1 |
| Waterloo | IA | 1.128 | 290 | 4 | 2.738 | 268 | 4 | 0.457 | -0.511 | -3.028 | 30 | 1 |
| Wausau | WI | 0.971 | 211 | 3 | 1.623 | 57 | 1 | 0.597 | -0.037 | -0.159 | 122 | 2 |
| Wenatchee | WA | 1.161 | 301 | 4 | 2.795 | 276 | 4 | 0.976 | 0.090 | 4.428 | 338 | 5 |
| West Palm Beach | FL | 1.113 | 285 | 4 | 2.902 | 301 | 4 | 0.911 | 0.156 | 3.084 | 289 | 4 |
| Wheeling | WV | 0.871 | 139 | 2 | 3.177 | 329 | 5 | 0.638 | -0.301 | 2.796 | 275 | 4 |
| Wichita | KS | 0.648 | 27 | 1 | 2.545 | 231 | 4 | 0.733 | 0.156 | 1.160 | 187 | 3 |
| Wichita Falls | TX | 0.806 | 98 | 2 | 1.804 | 91 | 2 | 0.697 | 0.159 | -3.458 | 21 | 1 |
| Williamsport | PA | 0.809 | 102 | 2 | 1.937 | 113 | 2 | 0.829 | 0.080 | -0.018 | 128 | 2 |
| Wilmington | NC | 1.269 | 323 | 5 | 2.255 | 108 | 2 | 0.796 | -0.061 | -1.902 | 55 | 1 |
| Wilmington | DE | 1.257 | 326 | 5 | 1.889 | 174 | 3 | 0.825 | 0.061 | 5.477 | 355 | 5 |
| Winchester | VA | 0.751 | 57 | 1 | 2.958 | 308 | 5 | 0.840 | 0.107 | 1.275 | 191 | 3 |
| Winston | NC | 0.880 | 149 | 2 | 1.630 | 60 | 1 | 0.857 | -0.075 | 3.410 | 306 | 4 |
| Worcester | MA | 1.376 | 351 | 5 | 2.561 | 234 | 4 | 0.970 | 0.314 | -0.160 | 121 | 2 |
| Yakima | WA | 1.156 | 299 | 4 | 2.159 | 157 | 3 | 0.914 | 0.286 | 4.177 | 333 | 5 |
| York | PA | 1.042 | 260 | 4 | 1.509 | 40 | 1 | 0.800 | 0.114 | -0.984 | 82 | 2 |
| Youngstown | OH | 0.821 | 106 | 2 | 1.533 | 44 | 1 | 0.957 | 0.369 | 0.937 | 177 | 3 |
| Yuba City | CA | 0.833 | 113 | 2 | 3.448 | 345 | 5 | 0.579 | 0.030 | 5.775 | 361 | 5 |
| Yuma | AZ | 0.862 | 133 | 2 | 2.808 | 279 | 4 | 0.873 | 0.252 | 5.885 | 362 | 5 |

Notes: Details for 3 integration measures (Mean, Sigma and R-Square Trend t-stat) are presented for all 384 MSAs. Mean is the average quarterly house price return. We compute house price returns for each MSA in our sample as the log quarterly difference in its FHFA repeat home sales price index. Sigma is the standard deviation of returns. R-Squares are the estimates of integration and are used to obtain R-Square trend t-statistics. R-Squares are obtained from fitting MSA returns to the factor model described in Appendix Table 1. The time trend t-statistics are estimated by regressing the R-squares for each MSA on a simple linear time trend for all available quarters of data. The final R-Squares pertain to 2010:Q1 for all 384 US MSAs. The change in R-Squares refers to the difference between estimates for 2010:Q1 and 1983:Q4 for each MSA. Each characteristic is ranked from lowest to highest in comparison to all 384 US MSAs. Each characteristic is also binned by quintile in comparison to all 384 US MSAs.